\documentclass[psfig]{aa}

\usepackage{graphicx}
\usepackage{natbib}
\usepackage{morefloats}
\usepackage{subfigure}

\begin{document}

\title{SPATIALLY--RESOLVED SPECTROPHOTOMETRIC ANALYSIS AND MODELLING OF THE
SUPERANTENNAE\thanks{Based on observations collected at the European Southern
Observatory, Chile, ESO No. 66.B-0706.}}

\author{Stefano Berta\inst{1},
Jacopo Fritz\inst{1},
Alberto Franceschini\inst{1}
Alessandro Bressan\inst{2}\\
\& Claudio Pernechele\inst{2} 
}

\institute{$^1$ Dipartimento di Astronomia, vicolo dell'Osservatorio 2, 35122 
Padova, Italy\\
$^2$ Osservatorio Astronomico di Padova, vicolo dell'Osservatorio 5,
35122 Padova, Italy\\
}

\offprints{Stefano Berta, \email{berta@pd.astro.it}}

\date{Received \today / Accepted .....}

\titlerunning{Optical Spectroscopy of the Superantennae}
\authorrunning{Berta S. et al. }

\abstract{We have performed spatially--resolved spectroscopy of the double--nucleated
Ultra-Luminous Infrared Galaxy  IRAS 19254--7245, ``the Superantennae'', along the
line connecting the two nuclei. These data are analysed with a spectral
synthesis code, to derive the star formation and extinction properties of the
galaxy. The star formation history (SFH) of the two nuclei
is similarly characterized by two different main episodes: a recent
burst, responsible of the observed emission lines, and an older
one, occurred roughly 1 Gyr ago. We tentatively associate this bimodal SFH
with a double encounter in the dynamical history of the merger. 
We have complemented our study with a detailed analysis of the broad band
spectral energy distribution of the Superantennae, built from published
photometry, providing the separate optical-to-mm SEDs of the two nuclei. Our 
analysis shows that: a) the southern nucleus is responsible for
about 80\% of the total infrared luminosity of the system, b) the L-band
luminosity in the southern nucleus is dominated by the emission from an obscured
AGN, providing about 40 to 50\% of the bolometric flux between
8 and 1000 $\mu$m; c) the northern nucleus does not show evidence for AGN
emission and appears to be in a post--starburst phase. 
As for the relative strengths of the AGN and starburst components, we find
that, while they are comparable at FIR and sub--mm wavelengths, in the radio
the Sy2 emission dominates by an order of magnitude the starburst.

}

\maketitle


\section{INTRODUCTION}

Starburst galaxies are characterized by a substantially enhanced ongoing star
formation  with respect to their past average. Though initially selected as having
intense optical emission lines, it has later become clear
that the main phases of star formation are highly attenuated in the optical
and luminous in the infrared (the IRAS-discovered Luminous and Ultra-Luminous 
Infrared Galaxies -- LIRGs and ULIRGs).
The Infrared Space Observatory (ISO), together with ground-based mm facilities 
such as SCUBA on the JCMT, showed the existence of distant galaxies with 
enhanced IR emission
(e.g. Elbaz et al. 1999, Barger et al. 2000, Smail et al. 2000, Franceschini 
et al. 2001), characterized by a strong cosmological evolution, and partly 
responsible for the cosmic IR background (Puget et al. 1996, Hauser at al. 1998).
This high-redshift population of relatively unfrequent but very luminous galaxies
constitute a new important cosmological component.
\begin{figure}[!ht]
\centering
\includegraphics[width=0.45\textwidth]{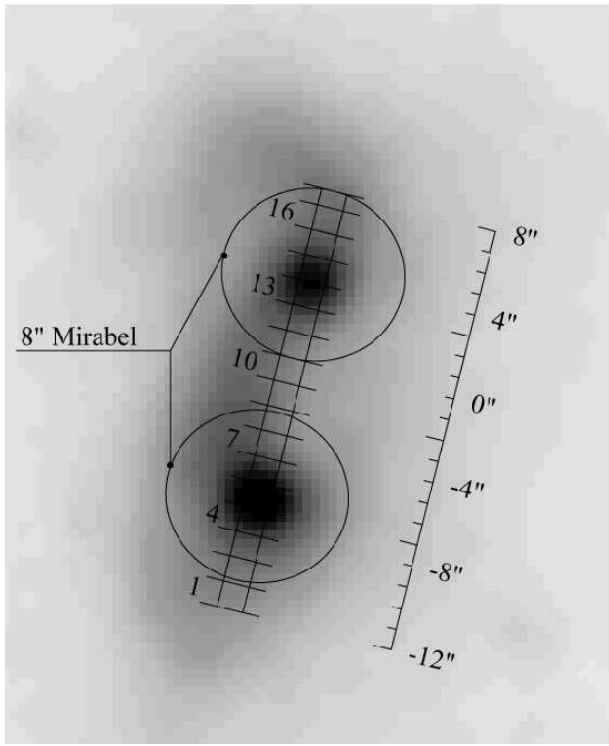}
\caption{Position of the central slitlet on the B image taken at the 3.6m
ESO telescope. The spatial spectral analysis has been obtained
by cutting the slit region in 17 slices, as numbered in the figure.
The two circular apertures centerd on each nuclei and with a diameter
of 8 arcsecs correspond to those adopted by 
\cite{mirabel1991}. See text for more details.
}
\label{fig:pos_fend}
\end{figure}

With luminosities in the range  10$^{11}$--10$^{13}$ L$_{\odot}$ and space 
densities similar to those of quasars (Soifer et al. 1986), the ULIRGs
are the most powerful objects in the nearby Universe \citep{SM96} and are considered 
to be the local counterparts of the newly discovered high-z IR luminous galaxies.
Understanding the nature of such ultra-luminous sources is then an important task
for observational cosmology, as they may correspond to the phases of formation, 
or the assembly, of spheroidal galaxies. 

Local infrared galaxies show evidence, from the optical emission lines, 
that their high luminosities are due to dust reprocessing of the
light emitted by young massive stars. However, the small sizes of the nuclear
star forming regions of compact ULIRGs are also consistent with a significant
fraction of their IR emission being provided by an AGN (Soifer et al. 2000).
The fact that in such objects the star formation rate (SFR) deduced from the IR
luminosity often exceeds that derived from the extinction-corrected optical 
line intensities (Poggianti et al. 2001) has been taken as an indication for
an obscured AGN contribution.
A concomitant starburst- and AGN-activity is sometimes clearly revealed by
broad optical and high excitation lines.

Whether this double--nature phenomenon is occasional or starburst activity is
systematically accompanied by an AGN, is still a matter of debate. 
The existence of a massive dark objects in the nuclei of galaxy spheroids
(Magorrian et al. 1998, Merritt \& Ferrarese 2001) may indeed suggest that
the AGN activity should have been a widespread phenomenon, at least in the past.

With the aim of further investigating the nature of luminous IR galaxies, we performed
a spatially--resolved spectro-polarimetric study of a small sample of ULIRGs 
selected from the \cite{gen98} mid--IR sample of 15 ULIRGs, based on a complete 
IRAS sample with $S_{60\mu m}\ge 5.4$ Jy (Pernechele et al., 2002). 
The far--infrared selection ensures minimal bias by dust absorption.
In this way we aim at identifying signatures in the SEDs of these sources 
which can be used to estimate the AGN contribution;
the mechanisms quenching out the central AGN after the dynamical destabilization 
of the gas; the relationship between the star formation process and the dynamical 
interaction.

In this paper we analyse the full-light spectrum of an ULIRG
prototype whose nature (staburst/AGN) is still intriguing: the Superantennae
(IRAS 19254--7245). 
At a redshift $z=0.061709$ \citep{strauss1992} and with a total infrared
luminosity of about 1.4$\times$10$^{12}$ L$_\odot$ (we use $H_0=65$ $[$Km
s$^{-1}$ Mpc$^{-1}]$), the Superantennae represents one
of the most powerful galaxy collisions in the nearby universe, exhibiting two giant
tails extending 350 kpc away from two nuclei, separated by $\sim$10 arcsec. 
This galaxy has been already studied by \cite{mirabel1991} and \cite{colina1991},
who report evidence of a type-2 AGN with a biconical galactic wind up to 800 km/s. 
\begin{figure*}[!ht]
\centering
\includegraphics{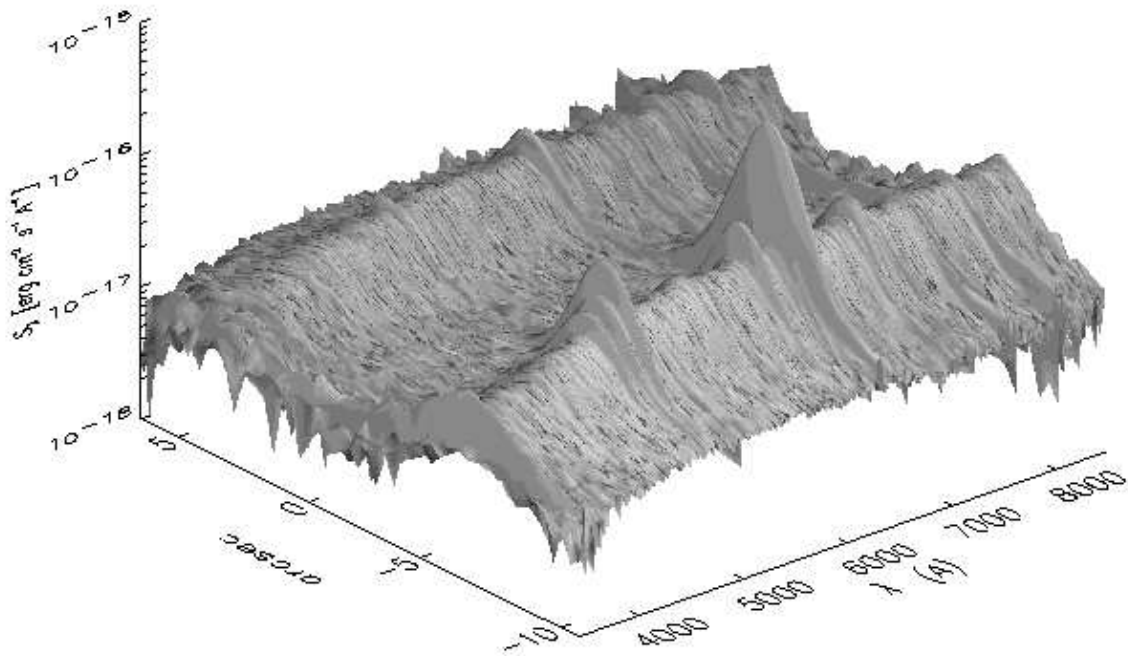}
\caption{The three--dimensional structure of the spectrum obtained with
the pipeline described in Section \ref{par:datared}. Different
emission lines are already clearly visible, H$\alpha$ standing over all
the others. The spatial axis corresponds to the one shown in Figure
\ref{fig:pos_fend}}
\label{fig:spettro_3d}
\end{figure*}

The paper is structured as follows. 
In Section {\bf \ref{par:datared}} we describe the observations and data reduction.
Section {\bf \ref{par:spat}} is devoted to the analysis of the most relevant
characteristics of the spatially--resolved spectra and of some physical properties, such
as extinction and star formation rate.
In Section {\bf \ref{par:fir}} we build a consistent broad band spectral energy
distribution of the two nuclei. These allow us to evaluate in Section 
{\bf \ref{par:sed}} the contribution of the star formation process
and the AGN to the bolometric infrared emission.
The optical spectrum of the two nuclei, corrected for the 
contribution of the AGN, is thus analyzed by means of a 
population synthesis tool described in Section {\bf \ref{par:modelli}},
the results are presented in Section {\bf \ref{par:results}}.
Section {\bf \ref{par:discussion}} is devoted to the discussion of the
star formation history  and the AGN non-thermal component of Superantennae.
Section {\bf \ref{par:Conclusions}} contains our conclusions.
We adopt $H_0=65$ [km s$^{-1}$ Mpc$^{-1}$].

\section{OBSERVATIONS AND DATA REDUCTION}\label{par:datared}

The spectropolarimetric observations of the Superantennae were done with EFOSC2 at
the 3.6m ESO telescope in La Silla, during the night of November the 1st, 2000.
 The seeing was about 1 arcsec and some veils were present
at low sky altitude. For this reason particular care has been 
taken in observing the target at its maximum elevation and 
chosing the standard star close to the object, 
in order to perform absolute flux calibration. We notice in advance
that, once properly corrected for aperture losses, the calibrated spectra
consistently fit the available literature optical and near--IR photometry (see
sect. \ref{par:sed}).

Polarimetric observations were obtained adopting a rotating $\lambda/2$
retarder plate
and a Wollastone prism, inserted before the grism. We choose the 236 {\rm
lines mm}$^{-1}$ grism, with a dispersion of 2.77 {\rm \AA\,
pixel}$^{-1}$, covering the 3685--9315 \AA \,wavelength range and a 
1.2$\times$20 {\rm arcsec} slit, corresponding to a spectral resolution of 21.2\AA.
The slit was oriented with a
P.A. of 14$^\circ$, such that both Superantennae's nuclei were included.

The data consist in a set of four couples of frames, obtained with retarder--plate
angles of 0, 22.5, 45 and 67.5 degrees, for a total integration time of 6000
seconds.
Johnson B and R band images were also obtained, with exposure times of 900 and 480
seconds, respectively.

The data reduction has been carried out with the use of {\sc iraf}'s standard tasks
and {\sc idl} procedures developed by ourselves.

Polarimetry is reported by \cite{pernechele2002}, while here we analyze the
full-light spatially-resolved spectrum.

The Wollastone prism splits the light into two perpendicularly polarized beams, 
which are projected onto the CCD. To avoid overlapping, the slit is transparent 
only in a series of 5 slots, separated by 5 opaque masks.
The prism causes each frame to appear as composed of 5 couples of spectra
(hereafter ``strips''), belonging to the 5 different slots.

The frames were bias and flat--field corrected in the usual way. Then the two
strips including the object's traces and two sky--strips were cropped.
Each strip was flux calibrated with the spectrophotometric standard star in the
centre of NGC 7293 \citep{oke1990} and then sky-subtracted.
Unfortunately, the interposition of the prism between the retarder plate and the
grism induces strong distortions of the frames, so we needed to provide a
distortion map by means of a polynomial fit using {\sc idl}.
Each spectrum was corrected for galactic reddening, 
assuming $R_V=3.1$ and E(B--V)=0.086 \citep{schlegel1998} 
and finally the various  strips, relative to 4 different polarization
angles, were combined together to produce the spatially
resolved total-light spectrum of Figure \ref{fig:spettro_3d}.

\begin{figure}[!ht]
\includegraphics[width=0.45\textwidth]{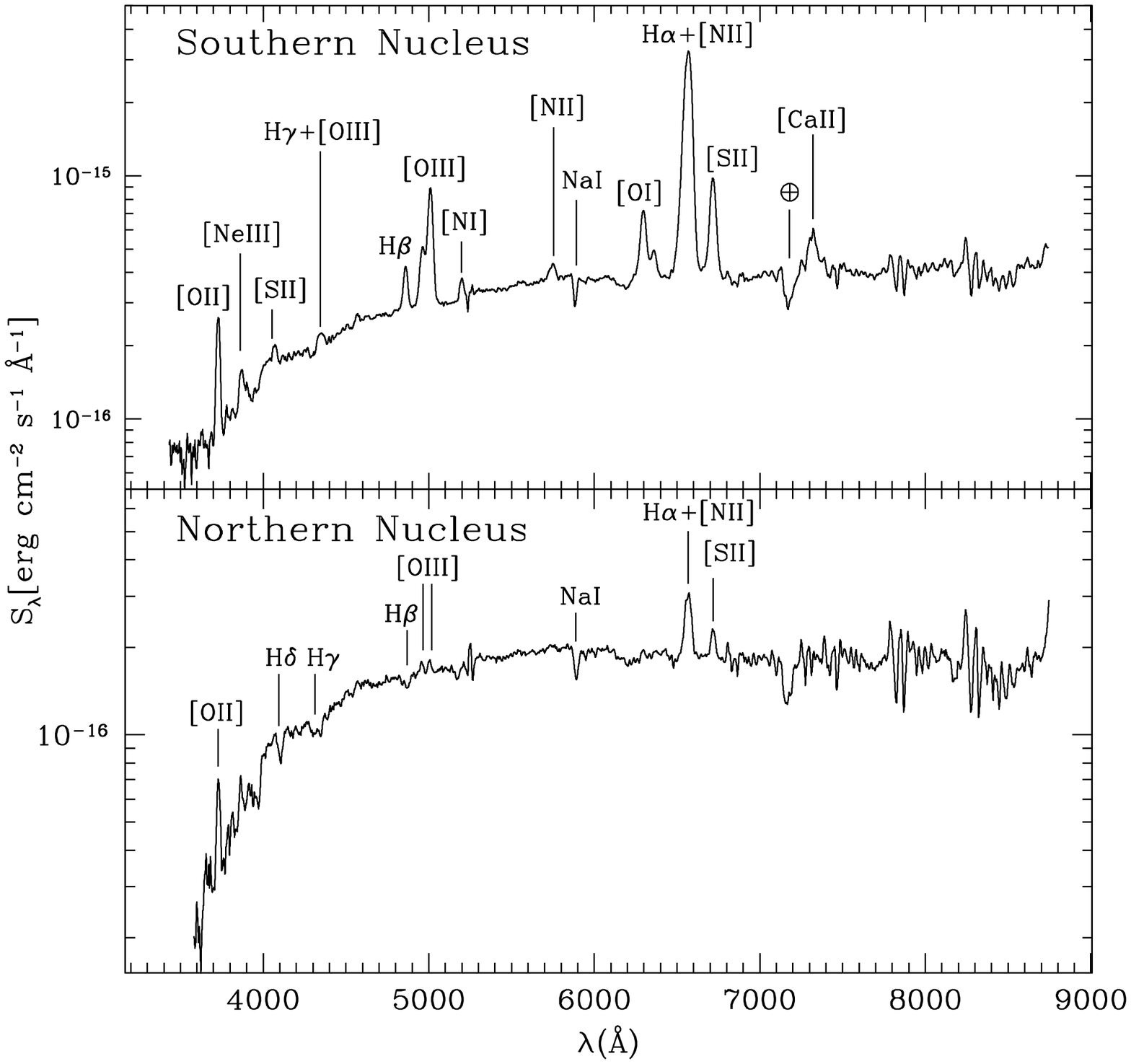}
\caption{The spectra of the two nuclei of IRAS 19254-7245 as extracted
from an aperture of $1.2''\times3.6''$ for the southern nucleus, and
$1.2''\times2.4''$ for the northern, respectively. 
A lot of emission lines typical of ULIRGs are recognizable.}
\label{fig:spettri_nuclei}
\end{figure}

The spectra of the nuclei were simply extracted through apertures of 1.2$\times$3.6
arcsec$^2$ (for the southern) and 1.2$\times$2.4 arcsec$^2$ 
(for the northern nucleus), see Figure \ref{fig:spettri_nuclei}. 
 At a redshift $z=0.06171$, 1 arcsec corresponds to $\sim1.4$ kpc.

\section{ANALYSIS OF THE SPATIAL DEPENDENCE OF THE MOST RELEVANT SPECTRAL FEATURES}\label{par:spat}

\begin{figure*}[!ht]
\begin{center}
\rotatebox{-90}{
\includegraphics[height=\textwidth]{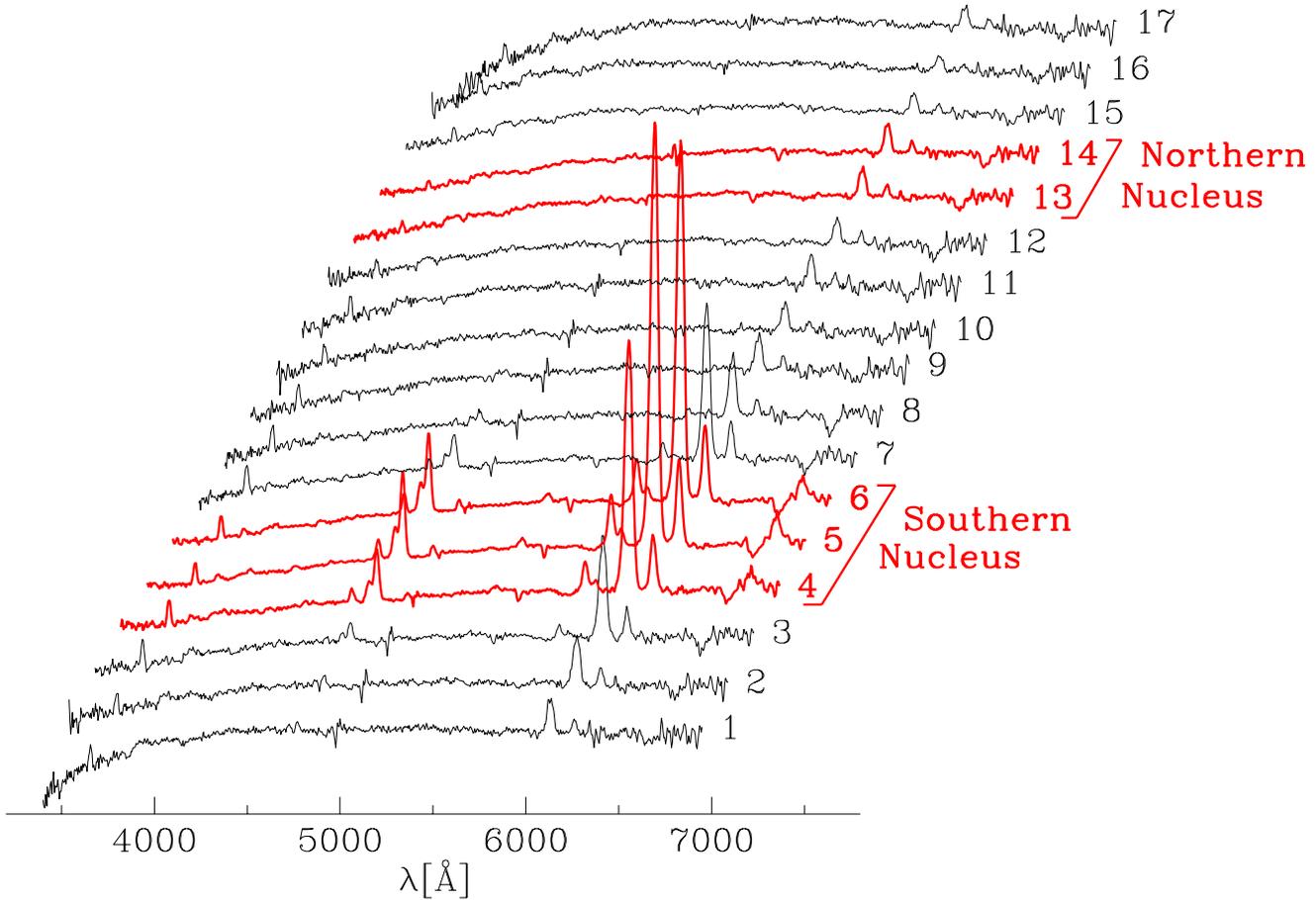}}
\caption{The seventeen different slices of Figure 
\ref{fig:pos_fend} are plotted in sequence, 
to show the trend of the continuum and other spectral features,
as a function of position along the slit.}
\label{fig:spettri17}
\end{center}
\end{figure*}
The high quality total--light spectrum 
allowed us to perform an analysis of the main spectral features of
Superantennae along the whole slit, hence to probe the spatial dependence of some
physical properties along the line connecting the two nuclei. 

To take into account the effects of the poor seeing, we binned our spectrum in 
steps of 3 pixels along the spatial axis (corresponding to 1.2 {\rm arcsec}) and 
thus obtained seventeen independent spectra corresponding to different regions 
of the galaxy.
In Figure \ref{fig:pos_fend} the $1.2''$ slit has been superposed on the B band 
image taken with EFOSC2: the different spectra obtained with the binning procedure
correspond to the regions numbered from 1 to 17 in the figure.
This also indicates the $8''$ apertures used by \cite{mirabel1991} in their
photometric study of the Superantennae (which we will use in the following
to estimate the contribution of the two nuclei to the total infrared flux).

\subsection{Emission lines}\label{par:profili}

\begin{figure*}[!ht]
\rotatebox{-90}{
\includegraphics[height=0.9\textwidth]{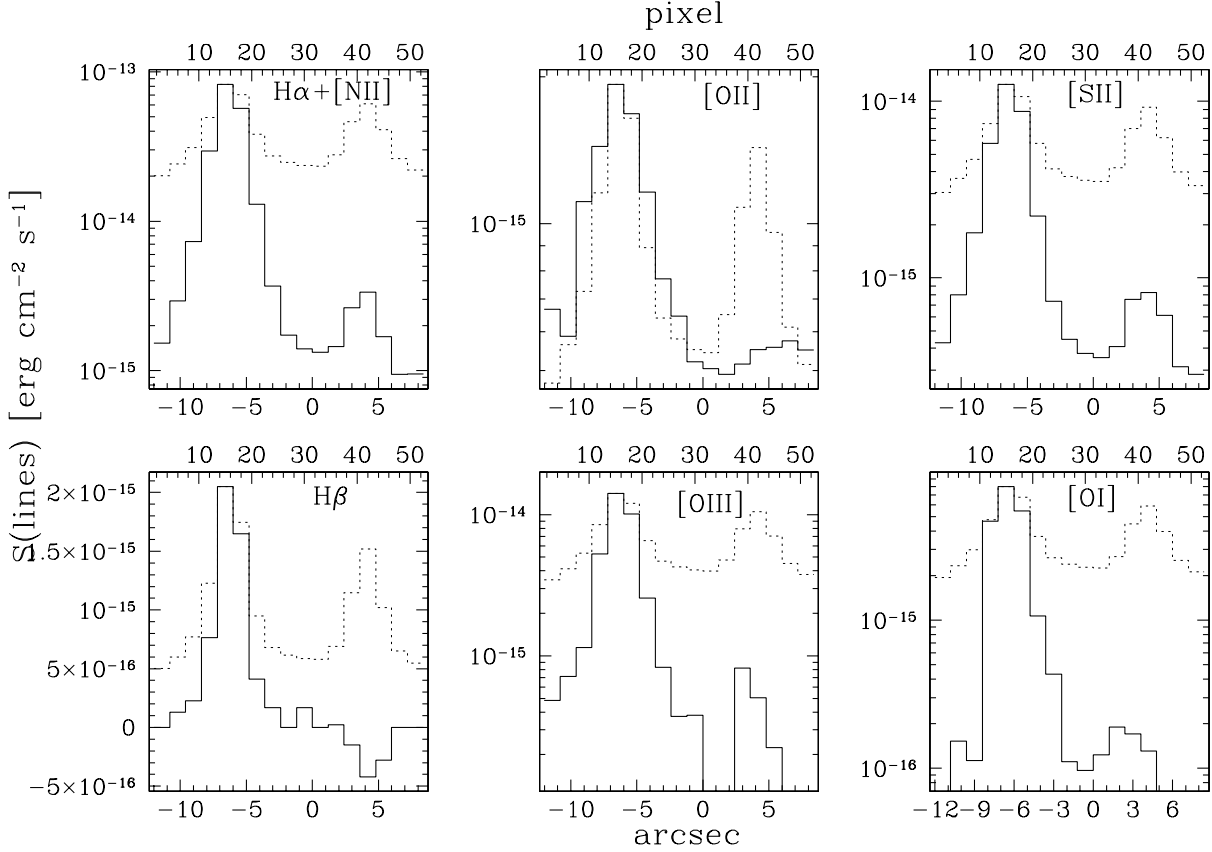}}
\caption{Spatial profile of the main emission lines. 
The solid histograms
represent the profiles of the various lines, as indicated in the
panels.  To study the spatial profile of the lines, we plot the continuum normalized to the peak value
of line emission in the sounthern nucleus (dotted histogram): going from the brightest nucleus to the outer regions, all the lines but {\sc [Oii]} fade more rapidly than the continuum.}
\label{fig:profili_6}
\end{figure*}

The 17 spectra corresponding to the different regions along the slit
(Fig.\ref{fig:pos_fend}) are shown in Figure \ref{fig:spettri17}. 
Several optical emission lines typical of starburst galaxies can be recognized 
(see also Fig.\ref{fig:spettri_nuclei}):
{\sc [Oii]}(3727 \AA), H$\beta$(4861 \AA), 
the sum of the two {\sc [Oiii]}(4959 and 5007 \AA), {\sc [Oi]}(6300 \AA),  
the blend of H$\alpha$ (6563 \AA) and the two {\sc [Nii]}'s (6548 and 6584 \AA), 
and the blend of {\sc [Sii]} (6717 and 6731 \AA).
 
The corresponding line fluxes are reported in Table 1.
Figure \ref{fig:profili_6} compares the profiles of the various line fluxes 
(solid line) along the spatial direction with the intensity of the continuum 
between 5500 and 5600 \AA\, (dotted line) normalized to the line intensity peak value.
It turns out that, with the exception of the {\sc [Oii]} (see Fritz et al., 2002, 
for a discussion), all other emission lines are much more peaked than 
the continuum, indicating that the ongoing star formation or the AGN activity is 
strongly concentrated within the nuclei.

The southern nucleus shows both narrow permitted and forbidden lines; 
we don't have enough spectral resolution to
separate  H$\alpha$ (6548 \AA) from the two {\sc [Nii]}'s (6563 and 6584 \AA),
but \cite{mirabel1991} reports that {\sc [Nii]} is stronger and sharper 
than H$\alpha$, thus revealing the Seyfert 2 nature of this emission. 

For the same reason we cannot
discriminate between {\sc [Oiii]}$\lambda$4363 and H$\gamma$ or {\sc
[Sii]}$\lambda$6717 and {\sc [Sii]}$\lambda$6731 and, consequently, 
we cannot derive an estimate of the electron density. 
By adopting the value of $N_e=670$ cm$^{-3}$ provided by \citep{colina1991}, 
we use the fluxes of {\sc [Nii]}$\lambda$5755 and {\sc
[Nii]}$\lambda\lambda$6548+6583 to estimate the electron density in the southern
nucleus (spectra \#4, 5 and 6). {\sc [Nii]}$\lambda$5755 was deblended from $[$Fe{\sc
vii]}$\lambda$5721, while the ratio between H$\alpha$ and {\sc
[Nii]}$\lambda\lambda$6548+6583 has been inferred from \cite{colina1991} data as
described next. The estimated electron temperature in the southern nucleus is 
$T_e \sim 14600$ K, 
consistent with the Colina et al. (1991) range of $10^4 - 2.5\; 10^4$ K.

The northern nucleus is characterized by a much lower star formation activity, as 
it is evident from the lower intensity of the lines or even the prevalence of
photospheric absorption for H$\beta$ and H$\gamma$.
A detailed analysis reveals that the {\sc [Oiii]}$\lambda$4363 + H$\gamma$ blend
is in emission in the three central spectra of the southern nucleus, possibly
because of a strong forbidden oxygen emission in the Seyfert component, but 
then moving outwards and to the northern nucleus it switches to absorption, 
dominated by photospheric H$\gamma$.
Also the blue edge of the continuum steepens significantly towards the northern 
component, showing that this nucleus is dominated by an older stellar population.

\begin{table*}[!ht]
\begin{center}
\footnotesize
\begin{tabular}{c r@{.}l r@{.}l r@{.}l r@{+}l  r@{.}l r@{.}l r@{.}l r@{.}l c}
\hline
\hline
Spectrum & \multicolumn{2}{c}{S({\sc[Oii]})} &   \multicolumn{2}{c}{S(B)} &     
\multicolumn{2}{c}{S(H$\beta$)} &         \multicolumn{2}{c}{S({\sc[Oiii]})}	  &       
    \multicolumn{2}{c}{S(V)} &
 \multicolumn{2}{c}{S({\sc[Oi]})}      &     \multicolumn{2}{c}{S(H$\alpha$+{\sc
 [Nii]})}	&    \multicolumn{2}{c}{S({\sc[Sii]})}  & E(B--V)  \\       
\hline
 1 &	6&68  &    0&28    &\multicolumn{2}{c}{--} &	   2.01&2.84   & 0&36 &\multicolumn{2}{c}{--} &  15&3 &  4&28	&   --        \\
 2 &	5&88  &    0&34    &	   1&28        &       2.47&4.69   & 0&43 &	   1&53      &   29&3	 &  8&03	& 0.429       \\
 3 &	11&1  &    0&41    &	  2&28         &       3.32&8.16   & 0&55 &	   4&97      &   72&9	 &  18&0	& 0.730       \\
 4 &	14&4  &    0&57    &	  7&65         &       16.1&36.7   & 0&88 &	   25&8      &   295&	 &  57&9	& 1.013       \\
 5 &	19&3  &    0&85    &	  20&5         &       42.3&99.3   & 1&47 &	   74&5       &   826&	 &  125&	& 1.090       \\
 6 &	16&8  &    0&75    &	  16&5         &       28.4&72.7   & 1&25 &	   46&8      &   570&	 &  87&8	& 0.974       \\
 7 &	11&6  &    0&43    &	  4&10         &       7.10&18.5   & 0&68 &	   7&13       &  130&	 &  22&4	& 0.827       \\
 8 &	77&0  &    0&31    &	  1&69         &       2.70&5.60   & 0&49 &	   3&20       &  36&7	 &  7&37	& 0.450       \\
 9 &	6&46  &    0&28    &\multicolumn{2}{c}{--} &	  0.814&2.91   & 0&44 &       1&31	  &  17&3    &  4&48	&   --        \\
10 &	5&21  &    0&28    &\multicolumn{2}{c}{--} &	  1.03&2.78    & 0&42 &       1&06	  &  14&0    &  3&73	&   --        \\
11 &	5&04  &    0&29    &\multicolumn{2}{c}{--} &\multicolumn{2}{c}{--} & 0&41 &	  \multicolumn{2}{c}{--}       &   13&3	 &  3&53&   --        \\
12 &	4&91  &    0&34    &\multicolumn{2}{c}{--} &\multicolumn{2}{c}{--} & 0&49 &	  1&55       &   14&5	 &  4&09&   --        \\
13 &	5&15  &    0&53    &	   -1&50       &       3.70&4.51   & 0&82 &	   1&75       &   26&3   &   7&54	&   --        \\
14 &	5&52  &    0&66    &	   -4&20       &       2.17&2.87   & 1&09 &	   1&31       &   33&5   &   8&26	&   --        \\
15 &	5&58  &    0&56    &	   -2&80       &       0.885&1.36  & 0&73 &\multicolumn{2}{c}{--} &   16&9   &   6&12	&   --        \\
16 &	5&75  &    0&42    &\multicolumn{2}{c}{--} &\multicolumn{2}{c}{--} & 0&47 &\multicolumn{2}{c}{--} &   9&44   &     3&13&   --  \\
17 &	5&51  &    0&30    &\multicolumn{2}{c}{--} &\multicolumn{2}{c}{--} & 0&39 &\multicolumn{2}{c}{--} &   9&52   &     2&84&   --  \\
\hline
South. Nucl. & 48&8	&    2&16    &	43&9             &	84.8&208     &  3&59 &	  156&          &  1680&   &   240&  	&  1.042 	 \\
North. Nucl. & 9&26	&    1&19    &  -6&60            &	3.73&5.03    &	1&91 &      4&03        &  57&0    &    13&2 	&   --   	 \\
\hline
\end{tabular}
\caption{Flux intensities of the main emission lines detected,  measured on the extracted spectra (no slit losses correction applied yet). 
The first column indicates the label of the spectrum as specified in 
Figure \ref{fig:pos_fend}.
In the next eight columns we report
the observed fluxes of {\sc[Oii]} (3727 \AA), H$\beta$
(4861 \AA), {\sc[Oiii]} (4959 + 5007 \AA), {\sc[Oi]} (6300 \AA),
H$\alpha$+{\sc [Nii]}  (6563, 6548 and 6584 \AA) and  {\sc[Sii]} (6717
\AA), respectively. Fluxes are in $10^{-16}$ [erg cm$^{-2}$ s$^{-1}$]. 
The last column shows the E(B--V) derived from
the Balmer decrement, as described in the text.  The last two rows report the fluxes for the two nuclei: the southern consists of spectra \#4, 5 and 6, the northern of \# 13 and 14. The nuclear fluxes are larger than the sum of fluxes in individual spectra, because they were measured on total nuclei extractions and not computed as sums.}
\normalsize
\end{center}
\label{linefluxes}
\end{table*}

\subsubsection{Extinction from the line ratios}\label{par:extinction}

The comparison of the observed ${\textrm H}\alpha/{\textrm H}\beta$
ratio with the theoretical
value of 2.85 predicted for a Case--B recombination \citep{osterbrock1989},
allows in principle to derive the average extinction, at least for the
southern nucleus where H$\beta$ is detected in emission.  To this end, 
we first need to quantify the contribution of the two {\sc [Nii]} lines at
6548 and 6584 \AA\ to the H$\alpha$+{\sc [Nii]} blend, because the Seyfert 2 nature 
of this nucleus prevents from adopting standard starburst values for the line ratio.
From the high resolution spectrum of \cite{colina1991}, we know that in
the southern nucleus $S({\textrm H}\alpha)=\frac{3}{4}S({\textrm {\sc [Nii]}
(6584)})$, hence assuming for the ratio between the two {\sc [Nii]} the the 
standard value ($1\over 3$), we get:
\begin{equation}\label{eq:rapporto_halpha}
S(\textrm{H}\alpha)=\frac{9}{25}S(\textrm{blend}).
\end{equation}  
Similarly, in the northern nucleus we obtain \citep{colina1991}
$S(\textrm{H}\alpha)\simeq\frac{1}{2}S(\textrm{blend})$.

Since Colina et al. (1991) spectra of the two nuclei were obtained through
a 3$''\times$6$''$ aperture, in the following we adopt eq.
\ref{eq:rapporto_halpha} on spectra \#2 to \#8, centered on the Southern nucleus
and the more standard value of $\frac{1}{2}$ in the Northern nucleus and outer
regions of the Superantennae.

To compare the empirical extinction estimate with the model prediction 
(see Section \ref{par:discussion}) we need to further
correct the observed H$\beta$ flux for the photospheric stellar absorption.
On this subject \cite{gonzalez} found that the contribution to the H$\beta$
equivalent width from
intermediate--age stars in starburst galaxies is $2-5$ \AA. Since they
do not account for an old disc population, which would decrease these values,
we have adopted a mean absorbing equivalent width of 2 \AA.
\begin{figure}[!ht]
\begin{center}
\includegraphics[width=0.45\textwidth]{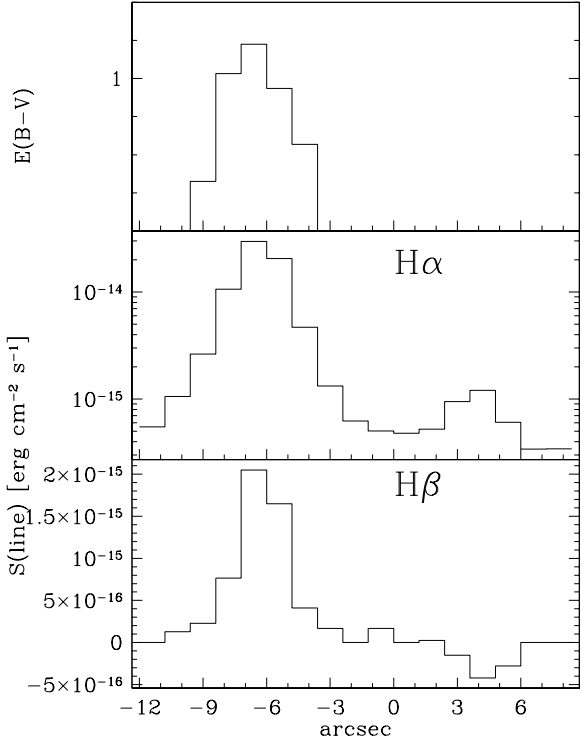}
\caption{Upper panel: spatial dependence of the extinction, E(B--V), 
as computed from the Balmer decrement along the slit. 
Middle and lower panels: H$\alpha$ and H$\beta$ fluxes. The H$\beta$ flux scale
is linear instead of logarithmic because the line is seen in absorption in the
Northern nucleus.}
\label{fig:extinction}
\end{center}
\end{figure}
The color excess E(B--V), derived from the H$\alpha$/H$\beta$ ratio is reported
in the last column of Table 1.   We plot in  Figure \ref{fig:extinction} the 
spatial dependence of the E(B--V) (upper panel) and of H$\alpha$ (middle panel) and
H$\beta$ line intensities.

The values of E(B--V)$\sim$1.0 {\rm mag} in the central regions obtained in this way 
are smaller than the value of E(B--V) $\simeq$1.4 {\rm mag} reported by 
\cite{colina1991}, but are consistent with those obtained by our
more detailed spectral analysis discussed in Sect. \ref{par:modelli},
which considers the presence of an old population. 
Also \cite{Vanzi2002} find $A_V=3.15$, consistent with our result.
This technique cannot be applied in the outer regions of the system and in the 
northern nucleus, where H$\beta$ is either not detected or in absorption. 
In those regions, spectral synthesis models (see fig. \ref{fig:ebv}) yield 
a mean value of E(B--V)$\sim0.4-0.5$, consistent with the typical average 
colour excess $\simeq$0.3 of spiral galaxies.

\begin{figure}[!ht]
\centering
\includegraphics[width=0.45\textwidth]{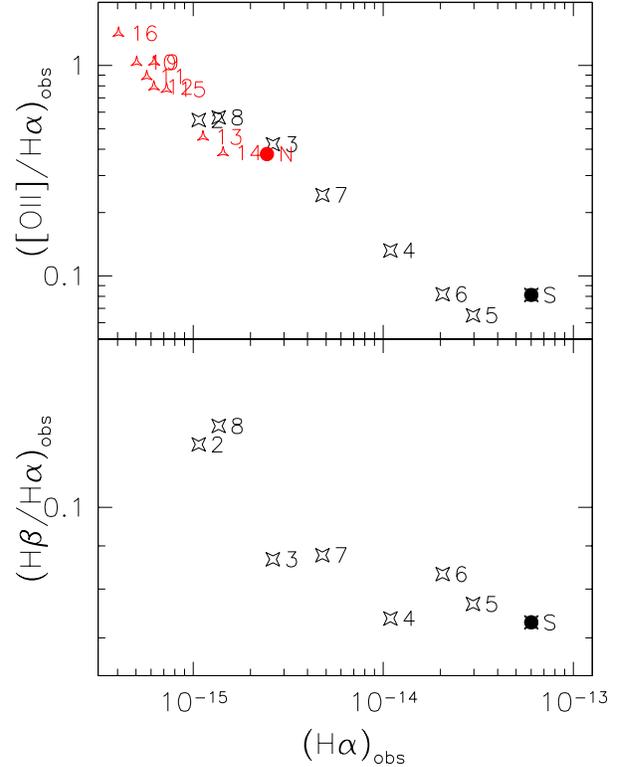}
\caption{{\sc[Oii]}(3727\AA)/H$\alpha$ and  H$\beta$/H$\alpha$ flux ratios
vs. H$\alpha$, before  extinction correction. Each data point is labelled
according to Figure \ref{fig:pos_fend} (the southern nucleus refers to spectra
\#4, 5, 6; the northern to \# 13 and 14). There is a clear dependence of the
ratios to the intensity of H$\alpha$, suggesting a tight relationship
between extinction and H$\alpha$ luminosity, indicative that the most
active (luminous) regions are also the more extinguished. 
See Fritz et al. (2003) for a detailed discussion.}
\label{fig:jans2}
\end{figure}

Hopkins et al (2001) and Sullivan et al. (2001) notice the
existence of a correlation between the extinction, derived from the
Balmer decrement, and the intensity of the H$\alpha$ emission line, in a
sample of star forming galaxies. The same trend is clear
in our spatially resolved analysis of Superantennae, when we plot the ratios of {\sc
[Oii]}(3727\AA) and H$\beta$ to H$\alpha$, as a function of the H$\alpha$
flux (Figure \ref{fig:jans2}). The observed ratio decreases as the H$\alpha$
intensity increases, confirming a strong correlation
between extinction and the star formation activity, as expected
if the latter is a function of the local density of gas and dust, and if
the extinction is mainly due to a foreground screen. 
The correlation involving {\sc [Oii]}(3727\AA), see top panel in Fig. 
\ref{fig:jans2}, seems to indicate that {\sc [Oii]} and H$\alpha$ emissions are
largely independent, in fact the {\sc [Oii]} strength only slightly depends
on position along the slit. 
We defer to Fritz et al. (2003) for a more detailed discussion of this effect.

\subsubsection{The Star Formation Rate}\label{par:sfr}

In star forming galaxies the extinction-corrected H$\alpha$ line intensity is a 
powerful tool to estimate the rate of ongoing star formation. 
Poggianti, Bressan \& Franceschini (2001), however, have warned that
for particularly extinguished objects the star formation rate determined
in such way may be significantly underestimated, by values up to a factor 3
compared with those derived from the far infrared luminosity.
Furthermore, as it will be discussed later, the Superantennae shows
strong evidence of a significant AGN component, probably contributing also to
optical emission lines.

With these caveats in mind, we have applied the relation of \cite{Kenni} to
the H$\alpha$ emission corrected for the {\sc [Nii]} contribution to the blend, 
 for extinction, and for the AGN contribution, 
to derive a preliminary estimate of the corresponding SFR, leaving a more 
accurate analysis to the following Sections.

Kennicutt assumes a Salpeter IMF between 0.1 and 100 M$_\odot$, 
solar abundance and a constant star formation rate during $10^8$ yr: 
\begin{equation}\label{eq:sfrHa}
SFR (M_\odot yr^{-1})=7.9\cdot10^{-42}L_{{\rm H}\alpha}\,({\textrm erg}
\,{\textrm s}^{-1}).
\label{eq2}
\end{equation}
The results are summarized in the first column of Table
\ref{tab:sfrmod} (see Section \ref{par:discussion}), with values ranging 
between $\sim 47\
M_\odot/yr$ in the southern nucleus and $\sim 0.4\ M_\odot/yr$ in the northern.

\section{THE BROAD BAND SPECTRAL ENERGY DISTRIBUTION}\label{par:fir}

Further information on the source comes from the analysis of the
broad band energy distribution. We have collected all the photometric
data available on IRAS 19254-7245 with the purpose of obtaining a
reliable spectral energy distribution for its two components.

\cite{mirabel1991} report magnitudes in the optical and near--IR, from the B
up to the L and N bands, within $8''$ apertures centered on the two nuclei, as well
as the total corrected IRAS fluxes at 12, 25, 60 and 100 $\mu$m with uncertainties
of 11\%, 5\%, 4\% and 8\%, respectively.
\cite{klaas2001} publish ISOPHOT measurements for the whole object at 10,
12, 15, 25, 60, 90, 120, 150, 180 and 200 $\mu$m and SEST data at 1300
$\mu$m. They quote uncertainties of about 5\% at 100--150 $\mu$m and up to
10\%--20\% for the other ISOPHOT passbands and for SEST.

\cite{charmandaris2002} report ISOCAM observations in various mid-IR bands
of the interacting system with a spatial resolution of few arcsec, 
hence resolving the two nuclei. 
These results are consistent with the photometric observations in the N-band by
Mirabel et al. for the southern nucleus, but are systematically fainter
for the northern one. 
Then we split the IRAS and the ISOPHOT long wavelength fluxes between the
two nuclei based on the Charmandaris et al. photometry: for $\lambda\le 12\ \mu$m 
we assumed a ratio between the IRAS and ISOPHOT fluxes belonging to the two nuclei equal 
to the ratio between the \cite{charmandaris2002} ISOCAM 11.4 $\mu$m fluxes; while 
longward 12 $\mu$m we adopted the ratio between the fluxes from the two nuclei by 
\cite{charmandaris2002} in the ISOCAM 15 $\mu$m filter (see table \ref{tab:fluxes_nuclei} 
and Fig. \ref{fig:sed_nord}).

Since the northern component is very faint compared to the
southern one, the uncertainties in method strongly affect the fluxes of the
northern nucleus. Therefore, as concerns the northern nucleus, we consider
fluxes obtained in this way reliable only shortward of 25 $\mu$m.
\begin{table}[!ht]
\begin{center}
\footnotesize
\begin{tabular}{c *{3}{c}}
\hline
\hline
		& South	8''& North 8''	& Global \\
\hline
B		&0.365 mJy&0.324 mJy& 1.401 mJy 	\\
V		&0.880 mJy&0.687 mJy& 2.997 mJy 	\\
R		&1.397 mJy&0.949 mJy& 4.220 mJy	\\
I		&2.052 mJy&1.407 mJy& 6.197 mJy	\\
J		&3.544 mJy&2.664 mJy&  --	\\
H		&5.720 mJy&3.353 mJy&  --	\\
K		&7.586 mJy&3.467 mJy&  --	\\
L		&20.79 mJy&2.117 mJy&  --	\\
N		&110.6 mJy&23.76 mJy&  --	\\
$S(12\mu m)$	&0.210$^*$  Jy&0.012$^*$  Jy& 0.222  Jy \\
$S(25\mu m)$	&1.217$^*$  Jy&0.025$^*$  Jy& 1.242  Jy \\
$S(60\mu m)$	&5.370$^*$  Jy& -- & 5.484  Jy \\
$S(100\mu m)$	&5.671$^*$  Jy& -- & 5.789  Jy \\
\hline
$S(6\mu m)$	& 0.0900 Jy & 0.0019 Jy & 0.0919 Jy \\
$S(6.75\mu m)$	& 0.1069 Jy & 0.0048 Jy & 0.1117 Jy \\
$S(7.75\mu m)$	& 0.1501 Jy & 0.0083 Jy & 0.1584 Jy \\
$S(9.62\mu m)$	& 0.0912 Jy & 0.0051 Jy & 0.0963 Jy \\
$S(11.4\mu m)$	& 0.1075 Jy & 0.0059 Jy	& 0.1134 Jy \\
$S(15\mu m)$	& 0.2840 Jy & 0.0059 Jy & 0.2899 Jy \\
\hline	
$S(10\mu m)$	& 0.116$^*$ Jy & 0.007$^*$ Jy & 0.123 Jy \\
$S(12\mu m)$	& 0.189$^*$ Jy & 0.011$^*$ Jy & 0.200 Jy \\
$S(15\mu m)$	& 0.392$^*$ Jy & 0.008$^*$ Jy & 0.400 Jy \\
$S(25\mu m)$	& 1.293$^*$ Jy & 0.027$^*$ Jy & 1.320 Jy \\
$S(60\mu m)$	& 5.456$^*$ Jy & -- & 5.570 Jy \\
$S(90\mu m)$	& 5.153$^*$ Jy & -- & 5.260 Jy \\
$S(120\mu m)$	& 4.183$^*$ Jy & -- & 4.270 Jy \\
$S(150\mu m)$	& 2.998$^*$ Jy & -- & 3.060 Jy \\
$S(180\mu m)$	& 2.253$^*$ Jy & -- & 2.300 Jy \\
$S(200\mu m)$	& 1.577$^*$ Jy & -- & 1.610 Jy \\
$S(1300\mu m)$	& 0.011$^*$ Jy & -- & 0.012 Jy \\
\hline
$S_{IR}$	&5.205$\cdot 10^{-13}$ &1.26$\cdot 10^{-14}$  &5.28$\cdot 10^{-13}$\\
\hline
\end{tabular}
\caption{Observed fluxes and magnitudes of the Superantennae. Optical and IRAS
data,  from \cite{mirabel1991}, are in the first thirteen rows, the next six rows report
the ISOCAM LW bands fluxes by \cite{charmandaris2002} and finally \cite{klaas2001} measurements are
in the last eleven rows; the 1300 $\mu$m datum was obtained with SEST, while the others
with ISOPHOT. Infrared observations marked with $^*$ have been splitted
between the two nuclei, by means of a procedure 
described in the text. Integrated IR fluxes
were obtained from IRAS data through \cite{SM96} calibration and are expressed in W$\cdot
$m$^{-2}$.}
\label{tab:fluxes_nuclei}
\normalsize
\end{center}
\end{table}

\begin{figure}[!ht]
\begin{center}
\includegraphics[width=0.45\textwidth]{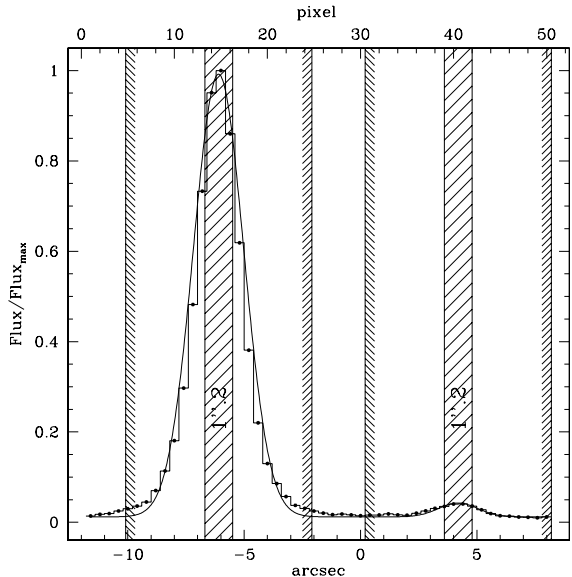}
\caption{Spatial profile along the slit, at the wavelength of H$\alpha$, used to
evaluate the amount of energy encircled within the $8''$  aperture centered on
each nucleus (delimited with shaded borders), which falls outside the slit 
($1.2''$ shaded regions). After performing a gaussian fit to the profile,
slit losses were estimated assuming that the  
bidimensional light profile is an axial-simmetric gaussian.}
\label{fig:losses}
\end{center}
\end{figure}

Almost 90\% of the infrared flux is emitted within the two $8''$ circular 
regions centered on the nuclei to which the optical photometry is referred, 
the southern nucleus being responsible for about $\sim 95\%$ of this flux.

\subsection{Slit-loss corrections}\label{par:losses}

To match the broad band SED obtained above with the optical spectra measured
within our slit, we have modelled the surface brightness profile of H$\alpha$ with
an axially-symmetric bidimensional gaussian. 
The free parameters, namely the peak value
and the variance $\sigma$, were estimated by fitting this
profile along the slit with a linear gaussian as shown in Figure \ref{fig:losses}. 
The values derived are $\sigma_{south}=
2.6$ pixels and $\sigma_{north}=2.1$ pixels, corresponding to $1.04''$ and
$0.84''$, for the southern and northern nuclei respectively.

The volume under the 3D gaussian is simply given by the integral of the profile
summed to the zero--level flux given by the continuum between the nuclei.
To evaluate  what fraction of the emitted flux falls outside the slit,
we integrated ``by slices'' within the 1.2$''\times$8$''$ apertures centered on
the nuclei. As a result, about the 58\% of the light coming from the
$8''$ aperture centered on the southern nucleus, and about 75\% of that
coming from the same region around the northern nucleus, were lost by our optical
spectroscopic observations.

The next step was to consider the amount of light enclosed in the
observed nuclear regions, namely within spectra numbered 4, 5 and 6
for the southern component, and 13, 14 for the northern one.
Considering that the observed H$\alpha$ emission is likely to appear
broadened because of the higher extinction in the very inner region of
the nucleus, we find that about $50\%$ of the flux observed from a
circular aperture of $8''$ on the southern object is emitted within this $1.2''
\times 3.6''$ aperture corresponding to spectra 4, 5, and 6. 
Assuming that the far-IR emission mimics the spatial
distribution the H$\alpha$ line, the ratio $S_{IR}/S_{5550}$ 
between the bolometric infrared and the V band fluxes turn out to be $\simeq 90$ 
for the southern nucleus 
\footnote{The flux at $5550$ \AA\, is computed as the mean value within about 
$50$ \AA\, centered at $\lambda=5550$}. 

Similarly we estimate a slit-loss of about $90\%$ from the northern nucleus
($1.2'' \times 2.4''$ aperture for spectra 13 and 14) and, correspondingly, 
a ratio $S_{IR} /S_{5550}\simeq$9.

\begin{figure*}[!ht]
\begin{center}
\rotatebox{-90}{
\includegraphics[height=\textwidth]{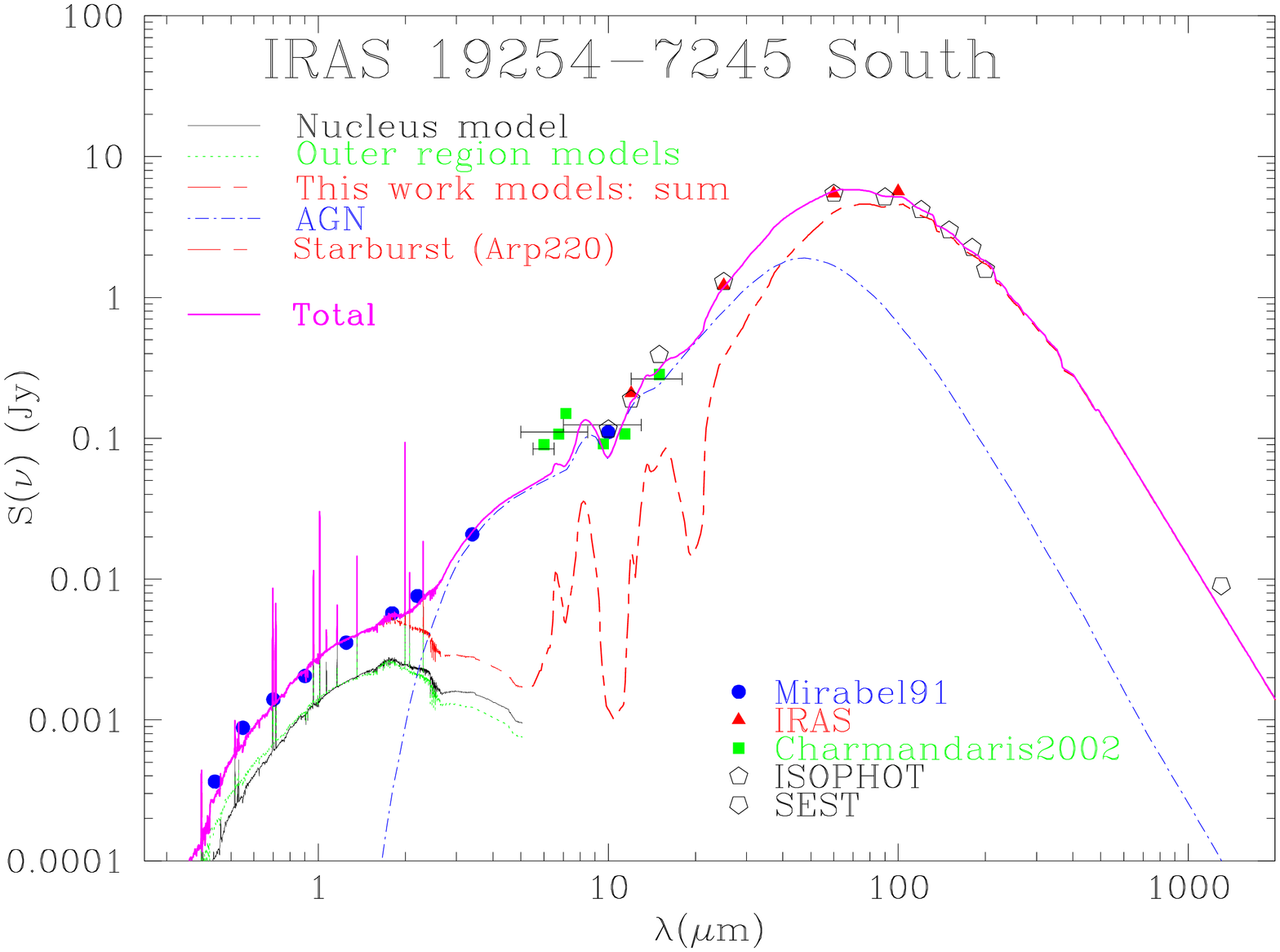}}
\caption{Spectral energy distribution of the southern nucleus. Data are from
\cite{mirabel1991}, \cite{laurent2000} and \cite{klaas2001} as described in the text.
The thick solid line is the fit to the global SED, obtained combining an optical--NIR
fit (dashed line short ward 5 $\mu$m) to our spectra (corrected to the $8''$
aperture) with an AGN model obtained with the code DUSTY (dot--dashed line) and a semiempirical starburst
Arp220--like template (dashed line long ward 5 $\mu$m). The thin solid and dotted lines
show the fit of the central and outer regions spectra. 
The N band width is slightly shifted upward with respect to the observed
datapoint, in order not to be confused with the LW2's.}
\label{fig:sed}
\end{center}
\end{figure*}

\section{THE SPECTRAL SYNTHESIS MODEL}\label{par:modelli}

The optical spectrum carries a wealth of information on the recent star formation 
history and dust extinction. In this Section we describe our detailed analysis of
the optical spectrum for each separate nuclear component.

As a best approach to describe the irregular star-formation histories of starburst
galaxies, we have adopted a completely {\sl free-form} galaxy spectral synthesis
model derived as a modofication of that discussed in \cite{PBF01}.
The optical observed spectrum is modelled as a combination 
of up to 10 simple stellar populations (SSP) of solar metallicity and different
age. Each SSP is meant to represent a formation episode of average constant  
star formation rate (SFR) over a suitable period $\Delta$t (see Table 
\ref{tab:risultati_modelli}).

As the main information provided by the spectrum are the shape of the
continuum, (nebular) emission lines and (photospheric) absorption lines,
we have chosen 3 classes of SSPs spectra to reproduce these features: young, intermediate and old.
The ages of the 10 populations have been chosen by considering the evolutionary
timescales of the stars associated with the observational constraints: 
the youngest generations
($10^6, 3 \cdot 10^6, 8 \cdot 10^6, 10^7$ yr) are responsible for the ionizing
photons producing the emission lines an for the UV continuum; 
 intermediate populations ($5 \cdot
10^7$, $10^8$, $3 \cdot 10^8$, $5 \cdot 10^8$, $10^9$ yr) have the strongest
Balmer absorption lines; while older generations of stars provide a 
significant contribution to the spectral continuum, 
hence affecting also the equivalent widths of the lines. The latters have been modelled
with a constant star formation rate (SFR) between 2 and 12 Gyr 
before the observation.

Each single population is assumed to be extinguished by dust in a uniform 
screen according to the standard extinction law
of the diffuse medium in our Galaxy [$R_V=A_V/E(B-V)=3.1$, Cardelli et al. 1989].
While a more complex picture of the extinction cannot be excluded, Poggianti et al.
have already shown that, in the case of obscured starbursts, the characteristics
of the emerging spectrum require a significant amount of {\sl foreground} dust
(screen model). For a uniform mixture of dust and stars, an increase of the 
obscuration does not yield a corresponding increase in the {\em
reddening} of the spectrum, and E(B--V) saturates to a value of $\sim 0.18$,
too low to account for the observed emission line ratios. In our case the
extinction value E(B--V) is allowed to vary from one stellar population to another
and the summed extinguished spectral energy distributions of all the single 
generations make up the total integrated spectrum.

The integrated spectra of the SSPs have been computed with
a Salpeter initial mass function (IMF) between 0.15 and 120 $M_{\odot}$.
We have adopted the Pickles (1998)
spectral library, after extending its atmospheres 
with Kurucz (1993) models outside its original range of wavelengths,
from 1150 \AA \ to 25000 \AA.
The composite (stars+gas) spectra have been obtained 
through the ionization code CLOUDY (Ferland, 1990).

The best-fit model was obtained by minimizing the differences between observed
and model selected features: the equivalent widths of five lines
([OII]$\lambda$3727,H$\delta$, H$\beta$ and H$\alpha$) and the relative intensities
of the continuum flux in a number of almost featureless spectral windows which
are chosen according to the spectra charachteristics.

As already anticipated our spectral resolution does not allow to
resolve the H$\alpha$-{\sc [Nii]} blend.
\cite{colina1991}, on high resolution spectra, have found that
H$\alpha$ is about $30\%$ fainter than {\sc [Nii]6584}. 
Because our spectral synthesis analysis does not model the {\sc [Nii]}{6548} line, 
we apply a correction factor to account for this contribution to
the blend.

A critical constraint that we impose to the model is to reproduce the observed 
infrared to optical flux ratio $S_{IR}/S_{5550}$.
Our predicted infrared flux is computed as the difference between the total
non-absorbed and the absorbed model spectra.

The minimization procedure applies the algorithm of Adaptive Simulated Annealing
\citep[ASA,][]{Ing1,ASA} to a suitable merit function (MF) based on the
differences between modelled and observed selected spectral features:
\begin{equation}\label{MF}
F=\sum_{i=1}^{n}\left(\frac{M_i-O_i}{E_i}\cdot W_i\right)^2
\end{equation}
where $M_i$ and $O_i$ are respectively model and observed fluxes (or
equivalent width), and  $E_i$ and $W_i$ are the corresponding errors and 
weights. The latter are used to reduce the weight of some observables particularly
difficult to model (as the OII line flux).

The minimization procedure checks if any of the SSPs older
than $10^7$ years contibutes by less than 1\% to the total flux at each wavelength.
This may happen, for instance, when extinction becomes too large; in that case
that SSP is no more relevant for the fit and is not used.
Obviously, the same does not apply to younger SSPs
because, though very extincted, their contribution to the FIR emission could not
be neglected.

Effectively only few populations dominate the sinthetic resulting spectrum: as
can be seen from table \ref{tab:risultati_modelli}, the number of SSP used in
the fits decreases to 4 or 5 tipically.

\section{THE AGN CONTRIBUTION TO THE BROAD BAND SPECTRAL ENERGY 
DISTRIBUTION}\label{par:sed}

The broad band SEDs of the two nuclei of the Superantennae, together with 
the best-fit optical synthetic spectra are shown in Figures 
\ref{fig:sed} and \ref{fig:sed_nord}, respectively.

\subsection{The southern nucleus}

The resulting SED of the southern nucleus is compared in Figure \ref{fig:sed}
with a model obtained from the combination of three different physical
components.

In the optical domain, it is the result of our spectral synthesis
analysis. The model is obtained combining two different fits, belonging to the
central (\# 4, 5, and 6 -- thin solid line in Figure \ref{fig:sed}) and outer
(\# 2, 3, 7 and 8 -- dotted line) regions spectra, normalized in order to
account for the slit losses. The two spectral combinations are used to
describe the nuclear and large-scale galactic emissions.
Indeed the sum of the two (short--long dashed line) reproduces
quite well the optical photometric data, as anticipated in section
\ref{par:datared}.
It is worth noticing that the spectral model
reproduces well also the photometric data beyond 7000 \AA \, up to the H band.

A first inspection shows that from this fit there remains a strong residual
excess in the K and L 
bands with respect to the stellar component fitting the optical data. 
Longwards of the H band, the observed fluxes seem to follow a power-law shape, 
possibly due to the Seyfert II emission, while the stellar spectral component
turns down.  This excess is even more evident in the L band, where the stellar
emission is fainter and dust reprocessing of young stellar emission is not yet
relevant. 
The observed J-L colour is about 3.9 mag, while that of the best fit galactic model 
is J-L$\simeq$2.7mag. The latter is consistent with the value observed 
in the northern nucleus, see Table \ref{tab:fluxes_nuclei} 
and Fig. \ref{fig:sed_nord}, which does not require an AGN contribution.

In order to reproduce this excess we added a QSO
template, requiring that the stellar plus AGN component, due to circumnuclear
dust, match the L band flux. 
\cite{pernechele2002} performed a detailed polarimetric analysis of the optical
emission of IRAS 19254--7245, revealing
a scattered polarized component at 2\% level only. 
For this reason we generate a totally obscured AGN model with the code DUSTY
(Ivezi\'c et al., 1999) assuming a spherical geometry and a standard ISM.
As an alternative, we also checked our fit using the totally obscured edge--on AGN models by
Andreani et al. (1999), finding a full consistence with the DUSTY results.

The free parameters for the AGN fit are the ratio $R$ between the inner and
outer radii of the spherical dust distribution and the optical depth
$\tau_\lambda$ at 0.55 $\mu$m.  

At wavelengths longer than 10 $\mu$m the contribution of the starburst starts
to rise again. We thus added an Arp220--like starburst template (short--long dashed
line) normalized in order to account for the remaining IR luminosity produced by the
southern nucleus. Our template was obtained by merging the {\sc grasil}
synthetic model by \citep{grasil98} of Arp220 to the CAM--CVF spectrum between 5
and 20 $\mu$m by \cite{charmandaris1999}.

We then imposed that the sum of the AGN and starburst templates should reproduce 
the observed optical-infrared SED. Our best fits for the AGN emission were obtained
for R values between 150 and 400, and an absorption $\tau_{0.55\mu {\rm
m}}=30-40$ mag. 
Adopting higher values of $\tau_{0.55\mu {\rm m}}$ would
underestimate progressively the K band and the L band fluxes, as $\tau$
grows, because of self absorption in the AGN dust emission.
By converse, lower extinction values would lead to an overestimation
of the NIR fluxes.

The parameter R, modelling the size of the circum-nuclear dust distribution
around the AGN, has a moderate impact on our global fitting procedure:
by increasing R we essentially add cold dust in the outer parts far away from
the central AGN, hence slightly increase the AGN emission around 100 $\mu$m.
However, there is limited variance allowed to this:
our best fit shows that with values R$<$150 or R$>$400 would completely spoil
the fit to the observed SED. 

In this way we estimate that the contribution of the AGN to the
bolometric IR flux (estimated from IRAS data)
ranges between about 40\% and 50\%. 

The resulting model is the thick solid line in Figure \ref{fig:sed}; we note
that it does not fit the observed 6$-$7 $\mu$m data, 
possibly characterized by strong PAH features. Actually DUSTY doesn't account
for the PAH emission and there could be also some residual PAH from the
starburst component, not reproduced by the Arp220 template. On the other hand
the cause of this discrepancy could also be the spherical geometry assumption.

This analysis, in turn, suggests that a similar fraction of
the H$\alpha$ emission could originate from the AGN. 
Indeed Cid Fernandes et al (2001), analysing the properties of a sample of 35
Seyfert II nuclei, find that it is possible to set some rough constraints on the
non--thermal H$\beta$ component from the properties of the {\sc [Oiii]}/H$\beta$
ratio: they show how values of {\sc [Oiii]}/H$\beta$ between 4 and 13 imply an AGN 
contribution to H$\beta$ between 30 and 100\%.
On this basis, the measured IRAS19254 South {\sc [Oiii]}/H$\beta$ of
$\sim 6$ leads to a non--thermal contribution of $\sim 40\%$ to the H$\beta$
emission, consistent with our estimate in the infrared. 

In the followings we will thus assume that about 40\% of the line emission 
is due to the AGN.

\subsection{The northern nucleus}

As far as the northern nucleus is concerned, we found essentially no evidence
for an AGN component. 
\begin{figure}[!ht]
\begin{center}
\includegraphics[width=0.46\textwidth]{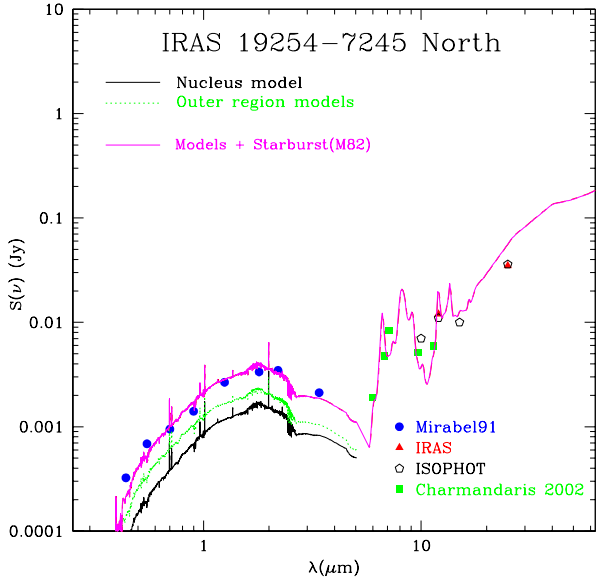}
\caption{The spectral energy distribution of the northern nucleus. The fit (thick
solid line) is obtained by adding an optical--NIR model short ward 5
$\mu$m, and a semiempirical starburst M82--like template (long ward 5 $\mu$m). 
The thin solid and dotted lines show the fit of the central and outer regions
spectra. No AGN component is needed in this case.}
\label{fig:sed_nord}
\end{center}
\end{figure}
The observed data consist in the set of optical and near--IR observations by
Mirabel et al. (1991), the mid--IR ISOCAM fluxes by Charmandaris et al. (2001)
and the IRAS and ISOPHOT fluxes up to 25 $\mu$m, obtained splitting literature
data for the whole (North + South) Superantennae through the method described in
section \ref{par:fir}. The latters are obviously more uncertain, but are still
consistent with Charmandaris et al. (2001) data, which are affected by a
$\sim 10- 20$\% uncertainty.

The fit is obtained by means of a pure stellar component,
modelled as in Section \ref{par:modelli}, from the optical to the NIR, 
and  a starburst component based on the prototype M82, at $\lambda > 5\mu$m.
The latter is obtained by merging a synthetic SED modelled with 
\cite{grasil98}, with the observed CVF spectrum by
\cite{schreiber2001} between 5 and 18 $\mu$m. 

The best fit is shown in Figure \ref{fig:sed_nord}. 
A small excess still present in the L band may be due to
PAH 3.3 $\mu$m emission not included in our spectral model.

\section{RESULTS OF THE SPECTRAL ANALYSIS}\label{par:results}
In this Section we discuss our results for the two nuclei separately and then we 
extend our analysis to the outer regions of the galaxy. 

The merit function (equation \ref{MF}) has been built on 10 continuum bands (3550$-$3670 \AA, 
3890$-$4020 \AA, 4140$-$4300 \AA, 
4400$-$4500 \AA, 4700$-$4820 \AA, 
5080$-$5180 \AA, 5360$-$5660 \AA,
5950$-$6110 \AA, 6880$-$6980 \AA \
and 7750$-$8500 \AA), 
and 5 spectral lines, H$\alpha+${\sc [Nii]}, H$\beta$ and {\sc [Oii]}3727, H$\delta$ 
and H$\gamma$. The latter two lines were assigned a weight of 0.3 because they are too
faint and, in the case of H$\gamma$, because of contamination by 
{\sc [Oiii]}4363, which is not considered in our models. 

We required that the model should reproduce the ratio $S_{IR}/S_{5550}$ between
the far-infrared and the optical flux, as defined in Sect. \ref{par:fir} and 
aperture corrected according to Sect. \ref{par:losses}.
The FIR flux was corrected for the contribution of the AGN, assumed to 
amount to 40\% of the bolometric IR emission (see Sect. \ref{par:sed}). 

The errors in the continuum flux were computed according to the local 
signal to noise ratio, while in the case of the lines fluxes the relative errors
were assumed proportional to $1/\sqrt{I_\lambda},$
where I$_\lambda$ is the flux emitted by the corresponding line.

Table \ref{tab:risultati_modelli} summarizes 
the parameters leading to the best fits: for each spectrum SFR and $A_V$ of the 
populations which build the sinthetic model are reported. 
Basically only 4 or 5 SSPs significantly contribute to the emitted spectra, therefore 
decreasing to 8 or 10 the number of parameters used to fit the 16 observed features chosen.

\subsection{Models for the southern nucleus}

The spectrum of the southern nucleus is obtained by summing the spectra with the 
strongest continuum and emission lines, namely those labelled with numbers 4, 5 
and 6 (see Fig. \ref{fig:pos_fend}). 

\begin{figure}[!ht]
\centering
\includegraphics[width=0.45\textwidth]{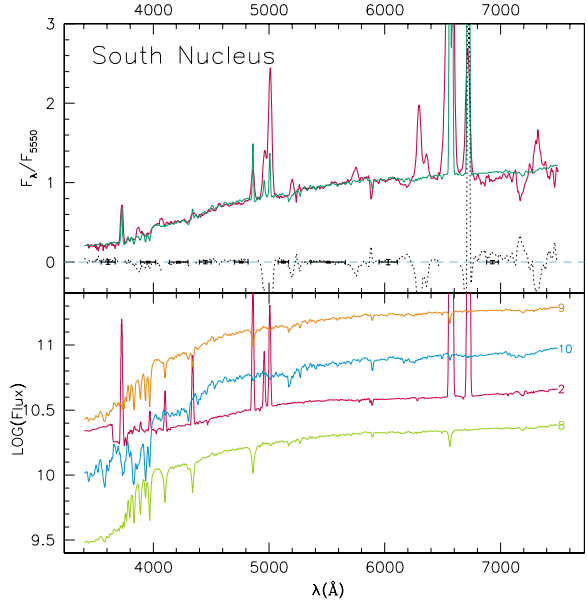}
\caption{Best fit model for the southern nucleus. Top panel reports the observed
(thick line) and the model (thin line) spectrum, the dotted line shows the
residuals between model and observed spectra. Bottom panel depicts the
contribution of each extinguished SSP (labeled from 1 to 10 with increasing
age). The intermediate age population dominates the continuum luminosity, while the
2-6 Myr population provides the emission lines.}
\label{fig:bf}
\end{figure}

Our best-fit solution is illustrated in Fig. \ref{fig:bf}. The upper panel shows
the observed and model spectrum (both normalized at 5500 \AA) and their
differences, the lower panel reports the contribution of the dominant stellar 
populations.
\begin{figure}[!ht]
\centering
\includegraphics[width=0.45\textwidth]{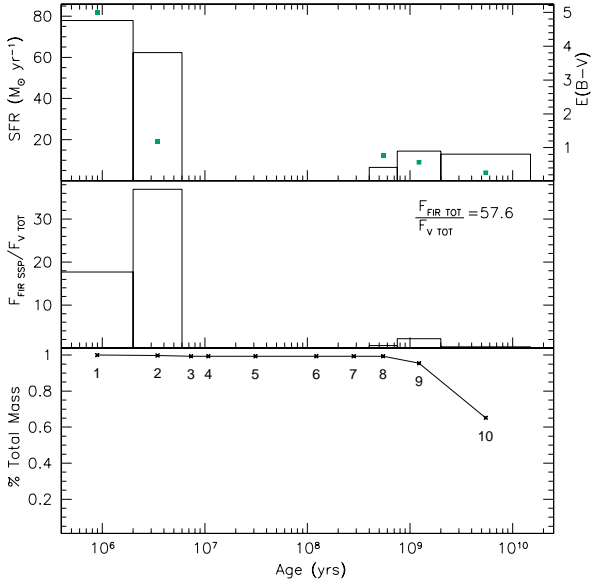}
\caption{SFR and extinction of the southern best fit model. Top panel: SFR
(histogram and left scale) and colour excess (filled squares and right
scale) for the different stellar populations. Middle panel: 
ratio between the bolometric FIR emission due to each SSP and the total V band flux. 
Notice that the sum over all populations is not 100, but turns out to be 57.6, 
which is the ratio between the {\em total} FIR and V fluxes.
This number is the conversion factor to obtain the differential contribution of 
each SSPs to the integrated infrared light. 
Bottom panel:
cumulative mass formed in the galaxy: about 35\% of the total stellar
mass has been processed by the intermediate age star formation episode, 
60\% belongs to the oldest. The youngest stars, significantly contributing to
the FIR bolometric emission and dominating the lines fluxes, provide almost no
contribution to the total luminous mass formed in the galaxy.
}
\label{fig:bf_m}
\end{figure}
\begin{figure}[!ht]
\centering
\includegraphics[width=0.45\textwidth]{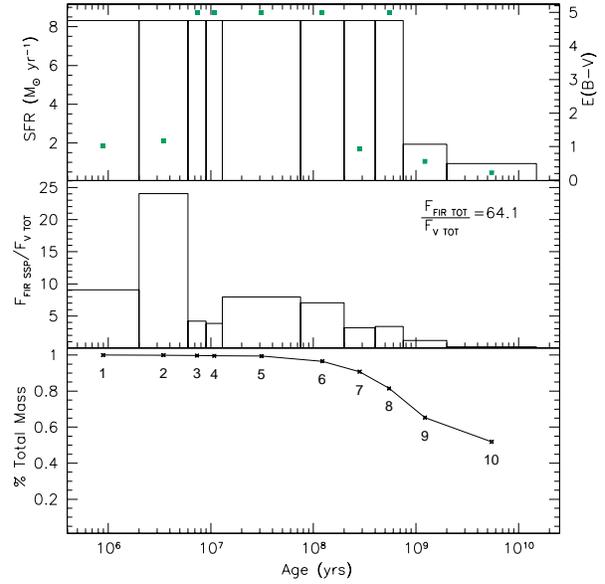}
\caption{A good fit to the optical spectrum of the southern nucleus (not
shown here) is found even by imposing a constant long burst of star
formation. However the fitting procedure "obscures"
unnecessary populations, which only contribute to the FIR. Due to this behaviour of the
color excess we exclude such a star formation scenario.} \label{fig:lbrst}
\end{figure}
All the modelled features are within $1\sigma$ error from their observed values,
except for the {\sc [Oii]} line ($3\sigma$) and for the band around 4760 \AA.
Fig. \ref{fig:bf_m} illustrates the star-formation history and extinction 
distribution of the best-fit model.  The optical spectrum is mainly contributed by
an old population (about 12 Gyr), providing about $65\%$ of the total
luminous mass, and an intermediate one (about $5\cdot10^8-10^9$ yr) forming about
35\% of the total luminous mass. On top of this there is the contribution of a 
strong and heavily extinguished burst of star formation, which account for
the strong observed emission lines and for the intense FIR flux (see Fig.
\ref{fig:bf}),  but give a negligible contribution to the total luminous
mass. The extinction shows the characteristic age-selective pattern
in which the youngest populations are more heavily obscured than the older ones.
In this case the youngest population is completely obscured with an 
E(B--V)$\simeq 5$ and cannot be seen in the optical spectrum.
The old populations with age $\ge 10^9$ yrs have E(B--V)$<$0.5.

In order to test the possibility of an alternative star formation history, 
characterized by a much longer starburst duration, we tried a constant star formation 
rate over the last $5 \cdot 10^8$ years. The corresponding star formation and
extinction patterns are shown in Fig. \ref{fig:lbrst}: with
this constraint, the code attempts to minimize the impact of the intermediate age
SSPs by increasing their extinction.
The predicted IR flux is within 3$\sigma$ from the observed value, and there are 
only little differences with respect to the best-fit model in the optical spectrum.
Although this solution cannot be formally excluded {\it a priori},
we consider it very unlikely, since extinction values so much higher for all the
intermediate age compared with the youngest populations do not have any physical 
meaning.

\subsection{Models for the northern nucleus}
\begin{figure}[!ht]
\centering
\includegraphics[width=0.45\textwidth]{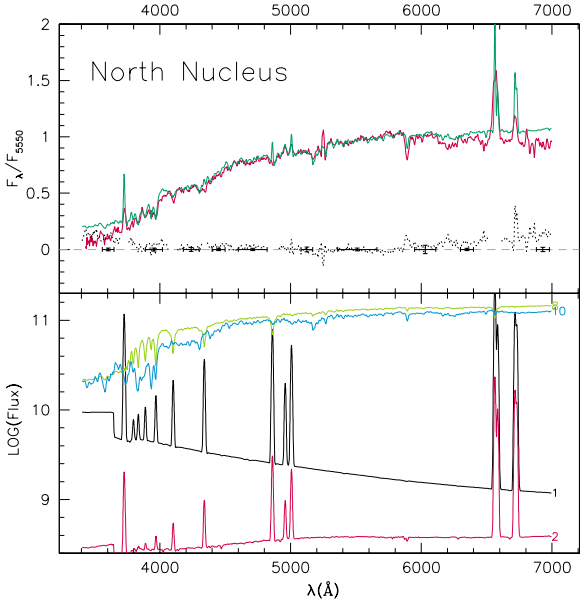}
\caption{Best fit to the northern nucleus spectrum. The mismatch
at the edges of the spectral range analysed is found in all other models we
have tried.}
\label{fig:n3}
\end{figure}
\begin{figure}[!ht]
\centering
\includegraphics[width=0.45\textwidth]{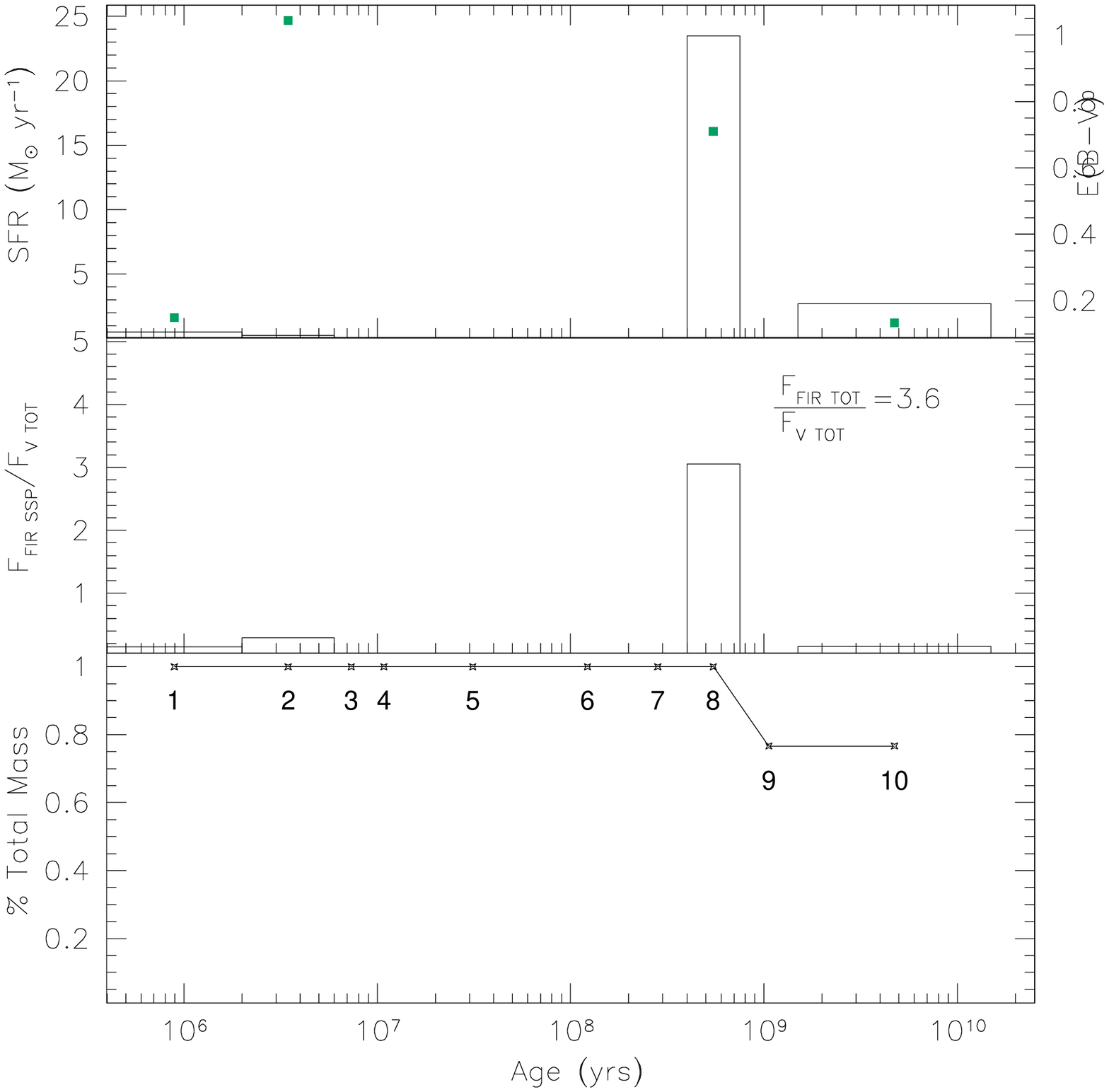}
\caption{SFR and colour excess of the northern nucleus best fit model. The
star formation history of this nucleus closely resembles that of the southern
nucleus, with two main star formation episodes, forming 75\% and 25\%
of the total stellar mass respectively. The intemediate age population
provides the bulk of the FIR flux. 
Extinction is significately lower than in the southern nucleus.}
\label{fig:n3m}
\end{figure}
The spectrum of the northern nucleus (spectral regions \# 13 and 14) was analyzed
with the same fitting procedure described previously. The results are 
shown in Figs. \ref{fig:n3} and \ref{fig:n3m}: the star formation history of 
this nucleus closely resembles that of the southern one, with two main star 
formation episodes and a recent, much weaker, burst. The extinction appears to 
be lower than that of the southern nucleus.
\begin{table*}[!ht]
\centering
\scriptsize
\begin{tabular}{c r r | c*{18}{c}}
\hline
\hline
\multicolumn{3}{c|}{SSP} 	&\multicolumn{18}{c}{Spectra}\\
\hline
\#& \multicolumn{1}{c}{Age} & \multicolumn{1}{c|}{$\Delta t$} &&\multicolumn{2}{c}{2}&&\multicolumn{2}{c}{3}&&\multicolumn{2}{c}{Nucleus
S}&&\multicolumn{2}{c}{7}&&\multicolumn{2}{c}{8}&&\multicolumn{2}{c}{9}\\
\cline{5-6}
\cline{8-9}
\cline{11-12}
\cline{14-15}
\cline{17-18}
\cline{20-21}
    	&		&	       && SFR  & $A_V$  && SFR  & $A_V$  && SFR     & $A_V$ &&  SFR   & $A_V$ && SFR  & $A_V$ && SFR & $A_V$ \\
1	&$10^6$		&$2\, 10^6$     && 0.09 &   --  && 12.14 &  5.27 && 77.85   & 15.25 &&  3.07  & 4.86 && 0.43 & 1.37 && 0.09 & 0.25 \\
2	&$3\, 10^6$	&$4\, 10^6$     && 0.21 &  2.27 &&  0.10 &   --  && 62.28   &  3.62 &&  1.89  & 2.04 && 0.02 &  --  && 0.05 & 2.28 \\
3	&$8\, 10^6$	&$3\, 10^6$     &&  --  &   --  &&   --  &   --  &&  --     &	--  &&   --   &  --  &&  --  &  --  &&  --  &  --  \\
4	&$10^7$		&$4\, 10^6$     &&  --  &   --  &&   --  &   --  &&  --     &	--  &&   --   &  --  &&  --  &  --  &&  --  &  --  \\
5	&$5\, 10^7$	&$6.2\, 10^7$   &&  --  &   --  &&   --  &   --  &&  --     &	--  &&   --   &  --  &&  --  &  --  &&  --  &  --  \\
6	&$10^8$		&$1.25\, 10^8$  &&  --  &   --  &&   --  &   --  &&  --     &	--  &&  1.32  & 2.04 &&  --  &  --  &&  --  &  --  \\
7	&$3\, 10^8$	&$2\, 10^8$     && 0.17 &  1.50 &&  0.25 &  1.35 &&  --     &	--  &&   --   &  --  &&  --  &  --  &&  --  &  --  \\
8	&$5\, 10^8$	&$3.5\, 10^8$   && 2.30 &  1.15 &&  2.18 &  1.35 &&  6.58   &  2.31 &&  1.05  & 1.52 && 2.26 & 1.52 &&  --  &  --  \\
9	&$10^9$		&$1.25\, 10^9$  &&  --  &   --  &&   --  &   --  && 14.48   &  1.73 &&   --   &  --  &&  --  &  --  && 1.45 & 1.46 \\
10	&$1.2\, 10^{10}$&$10^{10}$      && 0.49 &  1.03 &&  0.97 &  1.18 && 3.89    &  0.79 &&  1.72 & 1.52 && 1.15 & 1.43 && 0.33 & 0.75  \\
\hline
\multicolumn{3}{l|}{M$_{tot}$}&&\multicolumn{2}{c}{5.69e9}&& \multicolumn{2}{c}{1.05e10} && \multicolumn{2}{c}{5.97e10} && \multicolumn{2}{c}{1.77e10} && \multicolumn{2}{c}{1.23e10} && \multicolumn{2}{c}{5.12e9}\\
\multicolumn{3}{l|}{$A_V$(young)}
&&\multicolumn{2}{c}{1.741}&&\multicolumn{2}{c}{3.390}&&\multicolumn{2}{c}{3.838}&&\multicolumn{2}{c}{2.305}&&\multicolumn{2}{c}{0.905}&&\multicolumn{2}{c}{1.252}\\
\multicolumn{3}{l|}{$A_V$(tot)}
&&\multicolumn{2}{c}{1.169}&&\multicolumn{2}{c}{1.530}&&\multicolumn{2}{c}{2.278}&&\multicolumn{2}{c}{1.668}&&\multicolumn{2}{c}{1.437}&&\multicolumn{2}{c}{1.216}\\
\hline
\multicolumn{21}{c}{}\\
\multicolumn{21}{c}{}\\
\hline
& & &&\multicolumn{2}{c}{10}&&\multicolumn{2}{c}{11}&&\multicolumn{2}{c}{12}&&  \multicolumn{2}{c}{Nucleus
N}&&\multicolumn{2}{c}{15}&&\multicolumn{2}{c}{16}\\  
\cline{5-6}
\cline{8-9}
\cline{11-12}
\cline{14-15}
\cline{17-18}
\cline{20-21}
    &		    &		    && SFR   & $A_V$ && SFR   & $A_V$  && SFR	& $A_V$ && SFR   &$A_V$&& SFR   &$A_V$ &&  SFR   & $A_V$\\
1   &$10^6$	    &$2\, 10^6$     && 0.12  &  1.11 && 0.13  &  0.73  && 0.25  & 1.25  && 0.50  & 0.47&& 0.42  & 1.48 &&  0.06  & --   \\
2   &$3\, 10^6$     &$4\, 10^6$     && 0.03  &  1.11 &&  --   &   --   &&  --	&  --	&& 0.26  & 3.18&&  --   & --   &&  0.08  & 3.45 \\
3   &$8\, 10^6$     &$3\, 10^6$     &&  --   &   --  &&  --   &   --   &&  --	&  --	&&  --   & --  &&  --   & --   &&   --   &  --  \\
4   &$10^7$	    &$4\, 10^6$     &&  --   &   --  &&  --   &   --   &&  --	&  --	&&  --   & --  &&  --	& --   &&   --   &  --  \\
5   &$5\, 10^7$     &$6.2\, 10^7$   &&  --   &   --  &&  --   &   --   &&  --	&  --	&&  --   & --  &&  --	& --   &&   --   &  --  \\
6   &$10^8$	    &$1.25\, 10^8$  &&  --   &   --  &&  --   &   --   &&  --	&  --	&&  --   & --  &&  --	& --   &&   --   &  --  \\
7   &$3\, 10^8$     &$2\, 10^8$     &&  --   &   --  &&  --   &   --   &&  --	&  --	&&  --   & --  &&  --	& --   &&   --   &  --  \\
8   &$5\, 10^8$     &$3.5\, 10^8$   &&  --   &   --  &&  --   &   --   && 0.17  & 1.25  && 23.5  & 2.10&& 9.07 & 1.78 &&  0.62  & 0.40 \\
9   &$10^9$	    &$1.25\, 10^9$  && 0.50  &  1.11 && 0.36  &  0.73  && 1.94  & 1.25  &&  --   &  -- &&  --   &  --  &&  0.92  & 0.15 \\
10  &$1.2\, 10^{10}$&$10^{10}$      && 0.52  &  0.70 && 0.52  &  0.73  && 0.14  & 1.25  &&  2.70 & 0.37&& 0.22 & 0.40 &&   --  &  --   \\
\hline
\multicolumn{3}{l|}{M$_{tot}$}&&\multicolumn{2}{c}{5.84e9}&& \multicolumn{2}{c}{5.631e9} && \multicolumn{2}{c}{3.91e9} && \multicolumn{2}{c}{3.52e10}&& \multicolumn{2}{c}{5.37e9}&& \multicolumn{2}{c}{1.39e9}\\
\multicolumn{3}{l|}{$A_V$(young)} &&\multicolumn{2}{c}{1.101}&&\multicolumn{2}{c}{0.721}&&\multicolumn{2}{c}{1.233}&&\multicolumn{2}{c}{1.606}&&\multicolumn{2}{c}{1.467}&&\multicolumn{2}{c}{1.865}\\
\multicolumn{3}{l|}{$A_V$(tot)} &&\multicolumn{2}{c}{0.826}&&\multicolumn{2}{c}{0.721}&&\multicolumn{2}{c}{1.233}&&\multicolumn{2}{c}{1.633}&&\multicolumn{2}{c}{1.572}&&\multicolumn{2}{c}{0.639}\\
\hline
\end{tabular}
\footnotesize
\caption{Spatial dependence of the star formation rate (M$_\odot$/yr) and
extinction (A$_V$) in Superantennae, as derived
from the best fit models. Each row refers to
a different SSP; ages and durations, in years, are reported in
columns two and three respectively. The following pairs of columns refers to the
numbered galaxy regions and provide the SFR and extinction values for each SSP.
Values are not shown when the star formation rate is negligible. The bottom three
rows in each panel give the total stellar mass sampled by the spectrum and the 
A$_V$ extinction as computed combining the three youngest populations (young) and all the 10 SSPs (total). 
See text for more details.}
\label{tab:risultati_modelli}
\normalsize
\end{table*}

The old population contributes up to 75\% in mass and the intermediate age
population (0.5 Gyr) about 25\%. 

Note that there are problems in fitting the observed continuum
in the range 6000 \AA \,to 7000 \AA, which we were not able to solve,
even adopting different extinction laws, SSP metallicity, or different SSP 
combinations. A similar problem is found
between 3400 and 3800 \AA, where the model systematically overestimates the
observed spectrum, but here the observed spectrum has a low S/N ratio.

\subsection{Models for the regions outside the nuclei}

We have obtained best-fit models for the spectra of all the individual regions 
along the line connecting the two nuclei. Here both the continuum and line 
intensities are much lower than those from the two main nuclei 
(Fig. \ref{fig:profili_6}), and the star formation activity correspondingly low
(see Table \ref{tab:risultati_modelli}).
Nevertheless the main characteristic of the star formation 
history of the two nuclei, namely the evidence for an old stellar population and
for two main episodes of SF, one at $\sim 10^9\ yrs$ and one ongoing,
seems also reflected in the spectra of the extranuclear regions.

\section{DISCUSSION}\label{par:discussion}

We discuss here our previous results on the star formation history, 
extinction properties, and the AGN contribution in the {\sl Superantennae}.

\subsection{Spatial-dependent star formation and extinction in {\sl Superantennae}}

Table \ref{tab:risultati_modelli} reports our modelled values of the average
star formation rate (M$_\odot$/yr) along the slit for stellar populations at 
different ages. This quantity has been derived from the total stellar mass of
the population of a given age divided by the duration $\Delta t$ of the SF episod 
(third column in the table).
We have then computed the total stellar mass, sampled along the slit, and the 
corresponding visual extinction for the various populations.
\begin{figure}[!ht]
\centering
\includegraphics[width=0.45\textwidth]{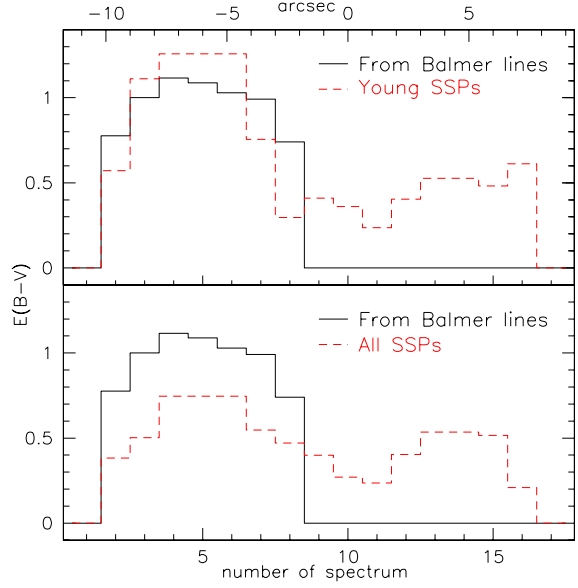}
\caption{The spatial dependence of the extinction, as
computed on the basis of the Balmer decrement  and through the 
spectrophotometric modelling of the spectra (see table \ref{tab:risultati_modelli}). 
The Balmer decrement has not been applied in regions 9 to 17
where H$\beta$ is not detected or appears in absorption.}
\label{fig:ebv}
\end{figure}

Figure \ref{fig:ebv} compares the E(B--V) values obtained with our global
fitting procedure with those derived from the Balmer decrement, after correcting
H$\beta$ for stellar absorption.  In the upper panel we show the E(B--V) distribution
obtained combining the three youngest SSPs, while
in the lower panel that derived averaging over all ten stellar populations. 
In the latter case the Balmer decrement technique appears to sistematically 
overestimate the extinction, while in the former there is a better agreement
at least close to the southern nucleus. This is because of the different extinction
properties of various populations, which the global fitting model can deal with,
but not the integrated spectrum.
Also, by accounting for the contributions of the intermediate age stars, 
the model provides a fair estimate of the reddening even in those regions 
where H$\beta$ is in absorption and the Balmer decrement technique is inapplicable
(e.g. spectra from 9 to 17).

Table \ref{tab:sfrmod} compares the {\em ongoing} SFR  
values obtained from our global fitting procedure (combining the two youngest SSPs)
with those derived by \cite{Kenni} calibrations:
\begin{equation}\label{eq:sfrK}
\begin{array}{lcr}
SFR\left( \textrm{\sc [Oii]} \right)&=&1.4\cdot10^{-41}\times L\left( \textrm{\sc [Oii]} \right)  M_{\odot} \cdot yr^{-1} \\
SFR\left( FIR \right)&=&4.5\cdot10^{-44}\times L\left( FIR \right)  M_{\odot} \cdot yr^{-1}
\end{array}
\end{equation}
and eq. \ref{eq2}. Before computing the SFR,
all the measured fluxes were corrected for internal absorption
using the extinction values found in Section 3. A value of E(B--V)$=$0.4, based on the
results of the spectral synthesis, was assigned to those spectra for which H$\beta$ 
is not detected or is found to be in absorption (see Section \ref{par:sfr}).
These relations refer to a Salpeter IMF between 0.1 M$_\odot$ and 100
M$_\odot$, while the IMF limits in our code are 0.15 and 120 M$_\odot$: the factor 
between the two is $\int_{0.1}^{100}m^{-1.35}dm/\int_{0.15}^{120}m^{-1.35}dm
\simeq1.162$.
The model's SFR of the southern nucleus exceeds by about 20\% the SFR estimate 
based on the FIR flux, but this is within the uncertainties of the calibrations 
if we considere that eq. (\ref{eq:sfrK}) refers to an average continuous SFR 
over the last 100 Myr, while the best-fit model in Figure \ref{fig:bf_m} assumes 
a very young short burst: the latter needs a higher SFR to supply a given FIR 
luminosity.

The {\sc [Oii]}$\lambda$3727 diagnostics leads to a higher SFR than the H$\alpha$, 
with a larger discrepancy in the circumnuclear regions. The {\sc [Oii]}$\lambda$3727
line has however a very different spatial profile from {\sc [Oiii]}$\lambda$5007,
H$\alpha$ or H$\beta$ (see Fig.\ref{fig:profili_6}), indicating a different degree
of ionization, possibly due to a metallicity gradient (see Fritz et al. 2002 for a
detailed discussion). Moreover 
the SFR obtained by the {\sc [Oii]}$\lambda$3727 line has a large intrinsic
uncertainty \citep{charlotl2001}.
A fraction of the star formation in the southern nucleus and in the circumnuclear
regions is apparently hidden (SFR from H$\alpha$ is smaller than SFR from FIR
even corrected for the AGN contribution), i.e. is not recoverable from the optical
spectrum.

Concerning the northern nucleus, the estimates based on the line fluxes and on the
spectrophotometric synthesis model analysis agree fairly well.
Apparently this region of the galaxy is likely in a post star-burst phase,
as suggested by the virtual absence of a very young component.
In this case, a substantial discrepancy between the SFR value based on the FIR
flux and those based on the other methods is apparent. 
Figure \ref{fig:n3m} indicates that the FIR
emission of the northern nucleus is produced mainly by a $5\cdot10^8$ years old
population; therefore applying the SFR calibration based on FIR 
leads to a misleading estimate, while our modelled SFR value appropriately
accounts for the ongoing star-formation activity. 
Because the youngest populations are those producing nebular emission lines, 
the H$\alpha$ and {\sc [Oii]} diagnostics are in much better agreement with 
our modellistic value.

Concerning the past star formation history, we find evidence for two
distinct episodes in this merging system. A recent, still ongoing burst producing the
optical emission lines and the strong infrared flux, and an older one which
took place about 1 Gyr ago. Remarkably enough, this double--peaked recent star 
formation history appears to be common to all the regions within our aperture.
Analysing a sample of interacting Ultraluminous IR Galaxies,
\cite{Murphy} suggest that the merging events may evolve through various
steps corresponding to different encounters, hence ULIRGs may show evidence
for multimodal stellar generations.
\cite{colina1991} give an estimate of about 1 Gyr for the age of the first encounter,
based on the extent and the speed of the gas in the tails of the system.
Our results are fully consistent with this overall picture of the merging process, 
and relate the two main episodes of activity with the first encounter
and the final merging of the two galaxies.
In this context, Figure \ref{fig:n3m} suggest that at the time of the second
enounter the gas in the northern nucleus was already almost exhausted.
During the first encounter about 30\% of the total stellar mass was formed,
while the remaining 70\% had already been formed by a pre-existing older
generation of stars.

Our model gives very similar masses for the two nuclei, consistent with the analysis
of \cite{colina1991} who, based on dynamical considerations, infer a ratio $\sim$1
between the masses of the two progenitor galaxies. 
\begin{table}[!ht]
\centering
\footnotesize
\begin{tabular}{c *{5}{c}}
\hline
\hline
Spectrum   & H$\alpha$ 	     &{\sc [Oii]}   &  FIR   & Model\\
\hline
1  &     --  	     &   --  	   &   --  &   	--   \\
2  &    0.22 	     &  0.59 	   &  0.25 &     0.21 \\
3  &    1.04 	     &  4.23 	   &  2.94 &     4.78 \\
South& 46.60 	     & 74.67 	   & 66.02 &   78.40 \\
7  &    2.28 	     &  6.79	   &  2.94 &    2.65 \\
8  &    0.29 	     &  0.85	   &  0.25 &    0.19 \\
9  &     --  	     &   --  	   &   --  &    0.07 \\
10 &     --  	     &   --  	   &   --  &    0.07 \\
11 &    0.11$^\ast$  &  0.13$^\ast$ &   --  &   0.05 \\
12 &    0.12$^\ast$  &  0.13$^\ast$ &   --  &   0.09 \\
North&  0.48$^\ast$  &  0.29$^\ast$ &  1.32 &   0.40 \\
15 &    0.14$^\ast$  &  0.15$^\ast$ &   --  &   0.16 \\
16 &    0.08$^\ast$  &  0.15$^\ast$ &	--  &   0.08 \\
17 &    0.08$^\ast$  &  0.14$^\ast$ &	--  &	 --  \\
\hline
\multicolumn{6}{l}{$^\ast$: values obtained assuming $E(B-V)=0.4$}\\
\multicolumn{6}{l}{\ \ \ \ in these regions of the galaxy.}\\
\end{tabular}
\caption{Values of the ongoing star formation rate SFR (in M$_\odot$/yr) derived from:
the main emission lines, after correction for extinction and for the AGN contribution,
and adopting the the Kennicutt's calibration; from the far-IR luminosity 
(Kennicutt's calibration); and from our global best-fit model. 
The latter has been computed combining
the two youngest single stellar populations. All SFRs dignostics adopt a Salpeter IMF 
between 0.1 and 100 M$_\odot$.
}
\label{tab:sfrmod}
\normalsize
\end{table}

\subsection{The contribution of the AGN}

Our analysis of the optical-FIR SED has found evidence for a substantial type-2
AGN component in the southern nucleus.  We have shown, in particular, that the 
J-L colour may be exploited as an estimator of the AGN contribution
to the MID and FIR emission. For example, while the northern nucleus has 
J-L$\simeq$2.7 value entirely consistent with a galaxy model dominated by evolved
stellar populations, the southern nucleus shows an excess in the L band, J-
L$\simeq$3.9 mag, that unequivocally proves the presence of an AGN. 
Fitting the global SED with a combination of an AGN and starburst templates, 
the contribution of the AGN has been constrained to be 40\%$-$50\% 
of the bolometric flux between 8 and 1000 $\mu$m. The uncertainty in this 
estimate is mostly due to uncertainties in the parameter $R$ setting the
scale size of the circum-nuclear dust distribution, which affects the dust
temperature distribution.

The optical spectrum also indicates the presence of a Seyfert-2 nucleus.
Pernechele et al (2002) provide evidence for $\simeq$2\% scattered polarized 
light around H$\alpha$ in the southern nucleus. 
Also the analysis of the {\sc [Oiii]}/H$\beta$ ratio
indicates a non--thermal contribution of $\sim 40\%$ at H$\beta$,
consistent with that required by the MID IR excess.

In X-rays, Pappa et al. (2000) showed that {\sl Superantennae} is
characterized by a very hard 2$-$10 keV spectrum with 
L$_X\simeq$2$\times$10$^{42}$erg/s, and Braito et al. (2002) found
evidence for a high-equivalent width FeI K$\alpha$ line typical of heavily
obscured type-2 AGNs.  

At radio wavelengths, the contribution of the AGN may be evaluated by comparing 
the observed flux with that expected from a starburst galaxy on the basis of the 
FIR/radio correlation (e.g. Sanders \& Mirabel, 1996):
\begin{equation}
q=\log \frac{\mbox{F}_{\mbox{FIR}}/(3.75\times 10^{12}\mbox{Hz})}{S_\nu
(1.49\mbox{GHz})/(\mbox{W m}^{-2}\mbox{Hz}^{-1})}\simeq 2.35\pm 0.15 ,
\label{qeq}
\end{equation}
where F$_{FIR}=1.26 \, 10^{-14}(2.58\, S_{60\mu{\rm m}}[Jy] +
S_{100\mu{\rm m}}[Jy])$ $\mbox{W m}^{-2}$.
Wright et al.(1994) report a radio emission of S$_\nu\simeq 102 mJy$ at 4.85 GHz. 
Assuming a radio slope $\alpha_R$=0.8,
typical of  starforming galaxies, the ratio in eq. (\ref{qeq}) becomes q$\simeq$2.71.
We find instead from Table \ref{tab:fluxes_nuclei} q$_{4.85\mbox{GHz}}\simeq$1.83,
significantly lower than the expected value for a normal starforming galaxy. 
Conversely, assuming that the starburst is responsible of $\sim$60\% of the
total infrared flux from Superantennae, and from eq. (\ref{qeq}), we expect
for a pure starburst component a 4.85 GHz flux of 7.8 mJy, more than one order 
of magnitude fainter than observed. So the AGN dominates the radio emission.

Finally, the two merging/interacting systems appear to have almost the same 
stellar mass ($\simeq$6$\times$10$^{10} $M$_\odot$ and 
$\simeq$3.4$\times$10$^{10}$M$_\odot$). From the correlation of the 
central Massive Dark Object (MDO) mass in galaxy nuclei with the bulge mass (Magorrian et al. 1998), we
would then expect that the two nuclei host similar-mass MDO's.
The fact that only the southern nucleus is site of AGN activity is consistent with
our discussed evidence that the northern nucleus has underwent a past 
activity episode (possibly involving both a starburst and an AGN), 
whose feedback and/or gas exhaustion have possibly quenched out any 
subsequent activity.

\section{CONCLUSIONS}\label{par:Conclusions}

We have presented a spatially--resolved study of the star formation and extinction
properties of the Ultra Luminous InfraRed Galaxy IRAS 19254--7245 (The Superantennae)
along the line connecting the two nuclei.

We have combined new optical spectroscopic observations with 
detailed broad band spectral energy distributions built from published photometry.
We have first attempted to quantify the presence of an obscured AGN 
-- for which there are evidences from both X-ray, optical and radio observations --
and its contribution to the source spectrum: we have found that the J-L colour
may be a powerful diagnostic tool. By means of an elaborate fit of the whole SEDs,
from the optical to the FIR, we have inferred that about 40--50\%
of the luminosity of the southern nucleus in the infrared is due to an
obscured AGN. The northern nucleus instead does not show any evidence of an AGN. 

After having accounted for the AGN contribution, the combination of the optical 
spectral data with infrared observations allows to disentangle between extinction,
age and star-formation effects: our spatially- and temporally-resolved analysis
is able to largely resolve all the degeneracies otherwise present in the integrated
spectra.   The data on Superantennae are consistent 
with a recent bimodal star formation history. The first episode of stellar
formation possibly corresponds to a first collision between the two progenitor
galaxies, about 1 Gyr ago, when a significant fraction 
of the gas mass was converted in stars (about 30\% of the present total
mass of the system). The merging onto the current configuration is due
to a more recent episode, adding a negligible fraction (few \%) of the total
stellar mass. 

We show that the northern nucleus has been characterized in the past
by a slightly older and less intense burst of star formation. 

The radio emission by {\sl Superantennae}, which is more than one order of magnitude
larger than expected from the starburst if we use the FIR/radio correlation,
confirms the presence of an AGN. Having said that, the radio flux cannot be
used to estimate the AGN contribution to the bolometric emission. Our analysis
also emphasizes that the technique used to estimate the redshift for 
dusty starbursts and based on the radio to mm flux ratio may provide highly
unreliable results if even a minor AGN component is present in the system.

\acknowledgements{We are grateful to G.L. Granato 
for providing us with models of SED of AGNs and to V. Braito 
for helpful discussions.
AB acknowledges support from the Italian Ministry
for University and Research (MURST) under grant Cofin 92001021149-002.}


\end{document}